\documentclass[12pt]{iopart}

\usepackage{epsf}
\usepackage{iopams}
\usepackage{amsfonts}
\usepackage{amssymb}
\usepackage{bm}
\usepackage{bbm}
\usepackage{nicefrac}

\usepackage[breaklinks=true,colorlinks=true,linkcolor=blue,urlcolor=blue,citecolor=blue]{hyperref}
\usepackage{graphicx}
\usepackage{epsfig}
\usepackage{latexsym}
\usepackage{dcolumn}
\usepackage{pifont}
\usepackage{siunitx}
\usepackage{cite}
\usepackage{epsfig}
\usepackage{latexsym}

\usepackage{dcolumn}
\usepackage{pifont}

\def\ii{{\mathrm{i}}}

\def\vdw{van der Waals}

\def\calH{{\mathcal H}}
\def\calL{{\mathcal L}}
\def\calF{{\mathcal F}}

\def\calV{{\mathcal V}}
\def\calW{{\mathcal W}}

\def\frakF{{\mathfrak F}}

\def\HFS{{\mathrm{HFS}}}
\newcommand{\fs}{{\mathrm{FS}}}

\def\LS{{\mathrm{LS}}}
\def\vdW{{\mathrm{vdW}}}

\def\DHF{{\calH}}
\def\DDD{{\calV}}
\def\WWW{{\calW}}
\def\FS4{{\calF}}

\newcommand{\lrgrule}{\rule[-4mm]{0mm}{11mm}}
\newcommand{\stdrule}{\rule[-2mm]{0mm}{6mm}}

\makeatletter

\newcommand{\Rmnum}[1]{\expandafter\@slowromancap\romannumeral #1@}

\newcolumntype{.}{D{x}{}{-1}}

\begin{document}

\title[{Pressure Shifts in High--Precision Hydrogen Spectroscopy: I.}]
{Pressure Shifts in 
High--Precision Hydrogen Spectroscopy: I. Long--Range 
Atom--Atom and Atom--Molecule Interactions}

\newcommand{\addrROLLA}{Department of Physics,
Missouri University of Science and Technology,
Rolla, Missouri 65409-0640, USA}

\newcommand{\addrCHEM}{Department of Chemistry,
Missouri University of Science and Technology,
Rolla, Missouri 65409, USA}

\newcommand{\addrHDphiltheo}{Institut f\"ur Theoretische Physik,
Universit\"{a}t Heidelberg,
Philosophenweg 16, 69120 Heidelberg, Germany}

\newcommand{\addrLEBEDEV}{P  N  Lebedev Physics
Institute, Leninsky prosp.~53, Moscow, 119991 Russia}

\newcommand{\addrMUC}{Max--Planck--Institut f\"ur
Quantenoptik, Hans--Kopfermann-Stra\ss{}e~1,
85748 Garching, Germany}

\newcommand{\addrRQC}{Russian Quantum Center,
Business-center ``Ural'', 100A Novaya street,
Skolkovo, Moscow, 143025 Russia}

\newcommand{\addrBUDKER}{Budker Institute of Nuclear Physics,
630090 Novosibirsk, Russia}

\author{U D Jentschura$^1$, C M Adhikari$^1$, 
R Dawes$^2$, A  Matveev$^{3,4}$\\
and N Kolachevsky$^{3,4,5}$}
\address{$^1$\addrROLLA}
\address{$^2$\addrCHEM}
\address{$^3$\addrLEBEDEV}
\address{$^4$\addrMUC}
\address{$^5$\addrRQC}

\begin{abstract}
We study the theoretical foundations for the 
pressure shifts in high-precision atomic beam 
spectrosopy of hydrogen, 
with a particular emphasis on transitions involving 
higher excited $P$ states.
In particular, 
the long-range interaction of an excited hydrogen atom in a $4P$ state with 
a ground-state and metastable hydrogen atom is studied,
with a full resolution of the hyperfine structure. 
It is found that the full inclusion of the 
$4P_{1/2}$ and $4P_{3/2}$ manifolds becomes necessary 
in order to obtain reliable theoretical predictions,
because the $1S$ ground state hyperfine frequency is 
commensurate with the $4P$ fine-structure splitting.
An even more complex problem is encountered in the 
case of the $4P$--$2S$ interaction, where the
inclusion of quasi-degenerate $4S$--$2P_{1/2}$ 
state becomes necessary in view of the dipole couplings
induced by the \vdw{} Hamiltonian.
Matrices of dimension up to $40$ have to be treated 
despite all efforts to reduce the problem to irreducible
submanifolds within the quasi-degenerate basis.
We focus on the phenomenologically important second-order 
\vdw{} shifts, proportional to $1/R^6$ where $R$ is the 
interatomic distance, and obtain results 
with full resolution of the hyperfine structure.
The magnitude of \vdw{} coefficients for hydrogen atom--atom
collisions involving excited $P$ states is
drastically enhanced due to energetic quasi-degeneracy;
we find no such enhancement for atom-molecule 
collisions involving atomic $nP$ states, 
even if the complex molecular spectrum
involving ro-vibrational levels requires a deeper analysis.
\end{abstract}

\vspace{2pc}

%
% Introduction
%
\section{Introduction}
\label{sec1}

Investigations of \vdw{} interactions 
involving excited states have attracted considerable 
attention~\cite{Ch1972,DeYo1973,AdEtAl2017vdWi,JeEtAl2017vdWii,%
PoTh1995,SaBuWeDu2006}.
In the retarded regime, the phase of an 
oscillation of a virtual transition changes 
appreciably over the time it takes light to 
cover the interatomic distance.
For excited reference states, 
one may obtain oscillatory long-range tails
from the energetically lower, virtual states,
which can give rise to interesting 
effects~\cite{SaKa2015,Be2015,MiRa2015,DoGuLa2015,Do2016,JeAdDe2017prl}.
We here analyze such interactions, with a particular
emphasis on the evaluation of the pressure shift
in the recent $2S$--$4P$ experiment
carried out in Garching~\cite{BeEtAl2017}.

In the presence of quasi-degenerate states,
the dominant contribution to the interaction is calculated 
by diagonalizing the Hamiltonian
matrix in a basis of 
quasi-degenerate states,
resulting in both first-order ($1/R^3$)
and second-order ($1/R^6$) energy shifts~\cite{JeEtAl2017vdWii}.
Using today's computer algebra~\cite{Wo1999},
it is possible to set up the calculation 
with hyperfine resolution,
i.e., to diagonalize the Hamiltonian 
matrix in a basis of states
where all hyperfine levels,
including their projections,
are resolved, resulting in rather large matrices.
In a quasi-degenerate basis, the energy separations
are on the order of the Lamb shift energy 
with virtual transition wavelengths in the 
centimeter regime; hence, the retarded regime
in this case is of no phenomenological relevance
because of the small absolute magnitude 
of the energy shift in this range.
In compensation, it is thus sufficient to 
treat the problem in the nonretardation approximation.

A significant motivation for an analysis
of the fine-structure, and hyperfine-structure resolved levels,
has been an ongoing experimental effort at a more
high-precision measurement of the hydrogen 
$2S$--$4P$ transition in Garching~\cite{BeEtAl2017},
where the resolution of the hyperfine structure,
together with the necessity to analyze collisional
frequency shifts, calls for a much improved theoretical
analysis of the \vdw{} interaction,
in comparison to previous approaches~\cite{JoEtAl2002},
which rely on nonrelativistic approximations.

As evident from the detailed analysis reported in the 
follow-up paper~\cite{MaEtAl2018jpb2},
excited-state interactions involving $4P$ states 
in contact with either ground-state 
$1S$ atoms or metastable $2S$ atoms are of prime 
importance~\cite{PoPriv2017,UdPriv2017}.
Phenomenologically, transitions to the $4P$ state
have been much more relevant than, say, transitions to 
$P$ states with $n=6$ (see Ref.~\cite{JeAd2017atoms}),
because of the much better accessible frequency range
of the transition for lasers (see Refs.~\cite{BeHiBo1995,BeEtAl2017}).
Specifically, $2S$--$4P$ measurements have been carried 
out by a number of groups~\cite{BeHiBo1995,BeEtAl2017},
whereas $2S$--$6P$ transitions
have not yet been measured to appreciable accuracy. The analysis is
sufficiently complex that either system could not be analyzed without the
use of computer algebra, due to the complex hyperfine structure state
manifolds. It is thus of prime importance
to generalize the treatment recently outlined in 
Ref.~\cite{JeAd2017atoms} to $4P$ states.
Furthermore, because of the possible presence of hydrogen 
molecules in any atomic beam undergoing
dissociation, it also becomes necessary to 
analyze the \vdw{} coefficients for atom--molecule 
collisions, 
even if we can anticipate that the 
\vdw{} coefficients will be drastically enhanced
for collisions involving only atoms, because of the 
quasi-degeneracy of excited states, which are removed
from each other only by the Lamb shift, fine- or 
hyperfine structure. Namely, the fine-structure and the hyperfine-structure 
splittings in the case of atom--atom interactions are very small compared 
to the energy differences between atomic and 
molecular quasi-degenerate levels, even if one consider 
possible excitations to ro-vibrational levels.
For example, in the case of the $4P(H)$--$1S(H)$ interaction,
the fine structure and the hyperfine structure splitting parameters 
are of the order of $2 \times 10^{-7} \,E_h$ and 
$9 \times 10^{-9}\,E_h$,  respectively, where $E_h = 27. 211396\,$eV 
is the Hartree energy~\cite{MoNeTa2016}. However,  
in the case of the $4P(H)$--$1S(H_2)$  interaction,  
the atom-molecules  degenerate states' separation 
is in the order of $2\times 10^{-2} E_h$ and the 
ro-vibrational level splitting is at-most $\sim 5.5\times 10^{-5} E_h$. 
The oscillator strengths, in either cases, are of the same order of magnitude. 
As the respective energy differences appear in the denominator 
of the propagator denominators within perturbation 
theory, which determine the $C_6$ coefficients, we
can anticipate that the 
so-called \vdw{} $C_6$ coefficients are enhanced
for atom-atom as compared to atom-molecule collisions.
This is explained in greater detail in Sec.~\ref{sec5}.

In order to understand the systems more deeply,
we should consider the particular properties of the 
\vdw{} Hamiltonian mediating the interaction.
Let us refer to the atoms participating in the 
interaction as atoms $A$ and $B$.
The static \vdw{} Hamiltonian (without retardation), in the dipole
approximation, involves the product of dipole 
operators of atoms $A$ and $B$.
An $SP$ state, with atom $A$ in an $S$ state
and atom $B$ in a $P$ state,
can be coupled, by the \vdw{} Hamiltonian,
to a state with atom $A$ in a $P$ state
and atom $B$ in an $S$ state.
Or, a state with atoms $A$ and $B$ in $S$ states,
can be coupled, by the \vdw{} Hamiltonian,
to a (possibly, quasi-degenerate) state with both atoms in $P$ states.
This implies that the \vdw{} interaction Hamiltonian
needs to be diagonalized in the 
energetically degenerate subspaces composed of the 
$SS$, $SP$, $PS$ and $PP$ states of the two atoms~\cite{JeEtAl2017vdWii}.
However, because of the usual dipole 
selection rules, the $SS$ and $PP$ manifolds 
do not mix with the $SP$ states,
and this reduces the size of the Hamiltonian 
matrices to be considered.
The latter fact can be verified explicitly on the 
basis of adjacency graphs which demonstrate the 
irreducibility of the matrices in the basis 
of the $SP$ and $PS$ states~\cite{AdDeJe2017aphb}.
Furthermore, interesting level crossings have been observed
in the two-atom interaction despite the 
irreducibility of the matrices~\cite{JeEtAl2017vdWii},
and an explanation in terms of higher-order 
interactions (distance within the adjacency 
graphs) has been described in Ref.~\cite{AdDeJe2017aphb}.

This paper is organized as follows.
In Sec.~\ref{sec2}, we outline the general formalism
behind our considerations (Sec.~\ref{sec2}),
before treating the $4P$--$1S$ interactions (Sec.~\ref{sec3})
and the $4P$--$2S$ interactions (Sec.~\ref{sec4}).
An interesting phenomenon is found in regard to 
the necessity of including both $4P_{1/2}$ as well as
$4P_{3/2}$ states into the basis,
and also $(4S; 2P_{1/2})$ quasi-degenerate virtual states.
We lay special emphasis onto the 
second-order \vdw{} shifts incurred by the 
levels, averaged over the magnetic quantum numbers,
as it is these numbers which are of highest phenomenological 
significance. 
Atom-molecule collisions are analyzed in Sec.~\ref{sec5},
and finally, conclusions are drawn in Sec.~\ref{sec6}.
SI mksA units are used here, except in Sec.~\ref{sec4},
where we switch to atomic units in order to keep 
formulas and mathematical expressions compact.

%
% General Formalism
%
\section{General Formalism}
\label{sec2}

%
% General Formalism
%
\subsection{Interaction Hamiltonian}
\label{sec2A}

Let us briefly review the derivation 
of the \vdw{} interaction 
and its application to excited states. 
Let $\vec x_A$ and $\vec x_B$ be the electron
coordinates, and $\vec R_A$ and $\vec R_B$ 
be the coordinates of the protons.
The total Coulomb interaction is
\begin{eqnarray} \label{eq:CoulombInt}
V_{\mathrm{C}} =
\frac{e^2}{4\pi\epsilon_0}
\left( \frac{1}{|\vec R_A - \vec R_B|}
+\frac{1}{\left|\vec x_A-\vec x_B\right|}
\right.
 \left.
-\frac{1}{|\vec x_A-\vec R_B|}
-\frac{1}{|\vec x_B - \vec R_A |} \right). \quad
\end{eqnarray}
One then uses the fact that the separation
$|\vec{R}_A-\vec{R}_B|$ between the two nuclei (protons) is
much larger than that between a given proton and its respective electron, that
is, much larger than both
$|\vec r_A| = |\vec x_A - \vec{R}_A|$ and
$|\vec r_B| = |\vec x_B - \vec{R}_B|$.
One then writes $\vec x_A-\vec R_B = \vec r_A + (\vec R_A - \vec R_B)$
and $\vec x_B-\vec R_A = \vec r_B + (\vec R_B - \vec R_A)$.
Expanding in $\vec r_A$ and $\vec r_B$, one obtains~\cite{Sa2016,Ad2017phd}
\begin{eqnarray}
\label{HVDW}
H_{\rm vdW} &=\; \frac{e^2}{4\pi\epsilon_0} \,
\frac{  \vec r_A \cdot \vec r_B - 3 \,
(\vec r_A \cdot \hat R) \,
(\vec r_B \cdot \hat R)}{R^3}\nonumber\\
&= \; \frac{e^2}{4\pi\epsilon_0 \, R^3} \,
\left( \delta_{k\ell} - 3 \, \hat R_k \hat R_\ell \right)
r_{Ak} \, r_{B\ell} 
= \;\frac{1}{4\pi\,\epsilon_0}\frac{ \beta_{ij} \, d_{Ai} \, d_{Bj} }{R^3} \,,
\end{eqnarray}
where $\vec R = \vec R_A - \vec R_B$, $R = | \vec R|$   
$\hat{R}=\vec{R}/R$ and   $d_{A}=e\, r_{A}$ is an electric dipole moment 
for atom $A$  and  $d_{B}$ is the same for atom $B$.
We have introduced the tensor
\begin{eqnarray}
\beta_{ik} = \delta_{ik} - 3 \frac{R_i \, R_k}{R^2} \,.
\end{eqnarray}
For definiteness, one chooses a quantization 
axis which enables one to resolve the magnetic projections in
the hyperfine manifolds. This motivates the choice
\begin{eqnarray}
\label{choice}
\vec R = R \, \hat e_z \,,
\end{eqnarray}
which is henceforth applied universally to all systems
studied in this paper. 

In our analysis of $4P$--$1S$ interactions, a   
typical virtual transition involving 
quasi-degenerate states would involve atom $A$ in a 
$|4P_J\rangle$ state (with $J = \frac12$ or $J = \frac32$), 
and atom $B$ still in the $|1S\rangle$ state.
This state is energetically degenerate 
with respect to a state where 
atom $A$ is in the $| 1S \rangle$ state,
and atom $B$ is in the $| 4P_J \rangle$ state.
Here, we further distinguish between 
absolute degeneracy (same unperturbed energy of the levels,
even including the hyperfine interaction),
and quasi-degeneracy, where levels are separated 
by the Lamb shift, or fine-structure interval.
For the absolutely degenerate case,
we incur first-order \vdw{} shifts,
linear in the \vdw{} Hamiltonian $H_{\rm vdW}$,
upon a rediagonalization of the total Hamiltonian.

An analogous situation is encountered 
for the $4P$--$2S$ interactions,
with the additional complication that 
an additional degeneracy exists with respect to 
virtual $(4S; 2P_{1/2})$ levels.
Namely, the lower $2S$ state is removed from the 
$2P_{1/2}$ state only by the classic Lamb shift,
and the $4S$ and $4P$ states are separated 
only by the $(n=4)$ fine-structure, or 
the $(n=4)$ Lamb shift.
Hence, additional virtual states have to be 
taken into account in the discussion of the 
$4P$--$2S$ interaction.

%
%
%
% Total Hamiltonian
%
\subsection{Total Hamiltonian}
\label{sec2B}

In order to evaluate the $4P$--$nS$ long-range interaction,
including hyperfine effects,
and fine-structure effects,
one needs to diagonalize the Hamiltonian
\begin{eqnarray}
\label{H}
H = H_{\LS,A} + H_{\LS,B} +
H_{\HFS,A} + H_{\HFS,B} + 
H_{\fs,A} + H_{\fs,B} + 
H_{\vdW} \,,
\end{eqnarray}
which sums over the atoms $A$ and $B$.
Here, $H_{\rm LS}$ is the Lamb shift Hamiltonian,
$H_{\fs}$ stands for the fine-structure splitting,
while $H_{\rm HFS}$ describes hyperfine effects.
We sum over the atoms $A$ and $B$.
The Hamiltonians are given as follows,
\numparts
\begin{eqnarray}
\label{HHFS}
H_{{\rm HFS}, i} =& \frac{\mu_0}{4\pi}\mu_B\mu_N\,
g_s \, g_p
\left[\frac{8\pi}{3}\vec{S}_i\cdot\vec{I}_i\,
\delta^{(3)}\left(\vec{r}_i\right)
\right.
\nonumber\\
&  \left.
+\frac{3 (\vec{S}_i\cdot\vec{r}_i) \,
(\vec{I}_i\cdot\vec{r}_i) -
\vec{S}_i\cdot\vec{I}_i \; \left|\vec{r}_i\right|^{\,2}}{\left|\vec{r}_i\right|^5} 
+ \frac{\vec{L}_i\cdot\vec{I}_i}{\left|\vec{r}_i\right|^3} \right] \,,
\\
\label{HLS}
H_{{\rm LS},i} =& \frac{4}{3}
\frac{\hbar^3 \, \alpha^2 }{m_e^2 \, c} \,
\left(\frac{\hbar}{m_e c}\right)^3\ln\left(\alpha^{-2}\right)
\delta^3\left(\vec{r}_i\right) \,,
\\
\label{HFS}
H_{{\rm FS},i} =&
- \frac{\vec p_i^{\,4}}{8 m_e^3 \, c^2} 
+ \frac{\pi \, \hbar^3 \, \alpha}{2 \, m_e^2 \, c} \, \delta^{(3)}(\vec r_i) 
+ \frac{g_s\, \hbar^2 \, \alpha}{4\, m_e^2 \, c \, \left|\vec{r}_i\right|^3}
\vec S_i \cdot \vec L_i \,,   
\\[0.0077ex]
\label{vdw}
H_{\rm vdW} =& \frac{e^2}{4 \pi \epsilon_0} \, 
\frac{x_A \, x_B + y_A \, y_B - 2 \, z_A \, z_B}{R^3}  \,.
\end{eqnarray}
\endnumparts
The fine-structure constant is denoted 
as $\alpha$, $\mu_0$ is the vacuum permeability,
and $m_e$ is the electron mass.
We treat the system in the non-recoil approximation.
The position and relative (with the respective nucleus)
momentum operators are $\vec{r}_i$ and $\vec{p}_i$,
while $\vec{L}_i$ is the orbital angular momentum operator.
Also, $\vec{S}_i = \vec\sigma_i/2$ is the (dimensionless) 
spin operator for the electron $i$, where $\vec\sigma$ is the vector 
of Pauli spin matrices,
and $\vec{I}_i$ is the spin operator for the nucleus
of atom $i$ (proton $i$).
According to Ref.~\cite{MoNeTa2016},
the protonic $g$ factors is $g_p\simeq \num{5.585695}$,
$\mu_B\simeq \SI{9.274010e-24}{\ampere\metre^2}$ is the Bohr magneton,
while $\mu_N\simeq \SI{5.050784e-27}{\ampere\metre^2}$ 
is the nuclear magneton. In order to simplify the 
expressions, we use the approximation $g_s = 2$ in the 
following calculations.

For the $4P$--$1S$ system, our convention is that 
the zero of the energy scale is the sum of the 
Dirac energies of the $1S$ and $4P_{1/2}$ states
(in the case of the $4P$--$1S$ interaction), 
and to the sum of the $2S$ and $4P_{1/2}$ states
(in the case of the $4P$--$2S$ interaction).
The zero point of the energy excludes 
both Lamb shift as well as hyperfine effects.
On the basis of the Welton 
approximation, we add the Lamb shift energy to 
the $S$ states, adjusted for the $S$--$P$ 
energy difference to match the experimentally
observed splitting, but leave the $P$ states untouched by
Lamb shift effects. Hence, strictly speaking, 
our definition of the zero point of the energy 
would correspond to the hyperfine centroid of the 
$|(4P_{1/2})_A (1S)_B \rangle$ states (see Sec.~\ref{sec3}),
and to the hyperfine centroid of the
$|(4P_{1/2})_A (2P_{1/2})_B \rangle$ states (see Sec.~\ref{sec4}).
The fine-structure energy is being added to the 
$4P_{3/2}$ states. For the calculation of the \vdw{} interaction 
energies, the precise definition of the 
zero point is not of relevance because only 
the energy difference in the quasi-degenerate basis matters.

The expression for
$H_{\rm LS}$ in Eq.~(\ref{HLS})
follows the Welton approximation~\cite{ItZu1980};
for the calculation of energy shifts, we shall replace
\numparts
\begin{eqnarray}
\label{defcalL}
\langle nS_{1/2} | H_{LS} | nS_{1/2} \rangle 
=& \; \frac{4\alpha}{3 \pi}\, \frac{\alpha^4}{n^3} \, m_e \,c^2 \, \ln( \alpha^{-2})
\to \calL_n \,,
\\[0.1133ex]
\langle nP_{1/2} | H_{LS} | nP_{1/2} \rangle =& \; 0 \,,
\end{eqnarray}
\endnumparts
where $\calL_n$ is the $nS$ Lamb shift,
which we understand as the 
$nS_{1/2}$--$nP_{1/2}$ energy difference.
These replacements are consistent with our
definition of the zero of the energy scale, 
as discussed above.
Throughout this paper, we perform final numerical
evaluations in the non-recoil approximation,
which corresponds to an infinite mass of the proton,
i.e., we set the reduced mass of the electron in hydrogen 
atom equal to the electron mass, and 
ignore the different reduced-mass dependence 
for the fine-structure and  the hyperfine-structure terms in the Hamiltonian.
Values for physical constants are taken from Ref.~\cite{MoNeTa2016}.

%
% Explicit Construction of the States
%
\subsection{Explicit Construction of the States}
\label{sec2C}

Even if the relevant procedure has recently 
been described in some detail in Sec.~IIB of 
Ref.~\cite{JeEtAl2017vdWii}, and in Sec.~I of 
Ref.~\cite{JeAd2017atoms}, we here recall how to 
construct the atomic states for the 
hyperfine-resolved $4P_{1/2}$--$1S$ interaction.
The relevant quantum numbers are
\numparts
\begin{eqnarray}
1S_{1/2}(F=0): n=1, \, \ell = 0, \, J =\frac12, \, F=0 \,, \\
1S_{1/2}(F=1): n=1, \, \ell = 0, \, J =\frac12, \, F=1 \,, \\
4P_{1/2}(F=0): n=4, \, \ell = 1, \, J =\frac12, \, F=0 \,, \\
4P_{1/2}(F=1): n=4, \, \ell = 1, \, J =\frac12, \, F=1 \,.
\end{eqnarray}
\endnumparts
Here, $n$ is the principal 
quantum number, while $\ell$, $J$, and $F$ are the 
electronic orbital angular momentum, the
total (orbital$+$spin) electronic angular momentum and the total
(electronic$+$protonic) atomic angular momentum, respectively.
Here and in the following, we denote by $F$ and $\frakF_z$ the 
total angular momenta (orbital$+$electron spin$+$nuclear spin)
of either atom $A$ or $B$, which can be specified 
for either atom by adding the respective subscript.
By contrast, $\frakF$ is their
vector sum $\vec\frakF = \vec F_A + \vec F_B$,
 so that, in particular, $\frakF_z = F_{z,A} + F_{z,B}$.

We denote by 
$\left\vert \pm\right\rangle_{e}$  the electron spin state, while
$|n,\ell,m\rangle_{e}$  denotes the Schr\"{o}dinger eigenstate (without spin). 
We need to add the nuclear (proton) spin $| \pm \rangle_{p}$
to the electron angular momentum.
For illustration, we indicate the 
explicit form of the hyperfine singlet $4P_{1/2}$ state,
\begin{eqnarray}\label{4p-singlet}
\fl \quad \left\vert n=4,\ell=1,J=\frac12, F=0,  F_{z}=0 \right\rangle 
= \frac{1}{\sqrt{3}} 
\left\vert +\right\rangle_{p} 
\left\vert +\right\rangle_{e} 
\left\vert 4,1,-1\right\rangle_{e}  
\nonumber\\
\fl \quad  \quad 
- \frac{1}{\sqrt{6}} 
\left\vert +\right\rangle_{p}
\left\vert -\right\rangle_{e} 
\left\vert 4,1,0\right\rangle_{e}
+ \frac{1}{\sqrt{3}} 
\left\vert -\right\rangle_{p} 
\left\vert -\right\rangle_{e}
\left\vert 4,1,1\right\rangle_{e} 
- \frac{1}{\sqrt{6}} 
\left\vert -\right\rangle_{p} 
\left\vert +\right\rangle_{e}
\left\vert 4,1,0\right\rangle_{e} \,,
\end{eqnarray}
while the hyperfine triplet states in the $4P_{1/2}$ manifold read as follows,
\begin{eqnarray}
\fl \quad \left\vert n=4, \ell=1,J=\frac12,F=1, F_{z}=0 \right\rangle
= -\frac{1}{\sqrt{3}}
\left\vert +\right\rangle_{p} 
\left\vert +\right\rangle_{e} 
\left\vert 4,1,-1\right\rangle_{e}  
\nonumber\\
\fl  \quad \;\; +\frac{1}{\sqrt{6}}
\left\vert +\right\rangle_{p} 
\left\vert -\right\rangle_{e} 
\left\vert 4,1,0\right\rangle_{e}  
+ \frac{1}{\sqrt{3}}
\left\vert -\right\rangle_{p} 
\left\vert -\right\rangle_{e}  
\left\vert 4,1,1\right\rangle_{e}  -
\frac{1}{\sqrt{6}} 
\left\vert -\right\rangle_{p}
\left\vert +\right\rangle_{e} 
\left\vert 4,1,0\right\rangle_{e}  \,,
\end{eqnarray}
and
\begin{eqnarray}\label{4Pstates}
\fl  \quad \left\vert n=4,\ell=1,J=\frac12, F=1, F_{z}=\pm1 \right\rangle \nonumber\\
\fl  \quad  \quad = \mp\frac{1}{\sqrt{3}} 
\left\vert \pm\right\rangle_{p} 
\left\{
\left\vert \pm\right\rangle_{e}  
\left\vert 4,1,0\right\rangle_{e} \right.
 -\sqrt{2}
\left.\left\vert \mp\right\rangle_{e}  \,
\left\vert 4,1,\pm1\right\rangle_{e}  \right\}.
\end{eqnarray}
\color{black}
Just like in Ref.~\cite{JeAd2017atoms},
we shall use the notation 
$| n,\ell, J, F, F_{z} \rangle$ for the 
thusly obtained states, using the vector 
coupling coefficients, with principal
quantum number $n$, orbital quantum number $\ell$,
total electron angular quantum number $J$,
total angular quantum number $F$ (electron$+$nucleus),
and total magnetic projection quantum number $F_{z}$.

Up to the hyperfine-fine-structure mixing term,
which is discussed in Eq.~(\ref{HFSFSmix}),
these states are eigenstates of the 
unperturbed Hamiltonian 
\begin{eqnarray}
\label{H0}
H_0 = 
H_{\LS,A} + H_{\LS,B} +
H_{\fs,A} + H_{\fs,B} +
H_{\HFS,A} + H_{\HFS,B} \,.
\end{eqnarray}
Based on the explicit representations 
of the relevant, hyperfine-resolved 
atomic states, one can easily develop a 
computer symbolic program, using computer algebra~\cite{Wo1999},
which determines the matrix elements of the 
total Hamiltonian~(\ref{H}) among all states 
within the hyperfine-resolved basis.
A different approach to the calculation of the 
matrix elements, especially useful for the evaluation of
matrix elements of the \vdw{} Hamiltonian,
is based on the Wigner-Eckhart theorem,
and will be discussed in the following.

%
% Wigner--Eckhart Theorem
%
\subsection{Wigner--Eckhart Theorem}
\label{sec2D}

It is very important and instructive to 
recall that the evaluation of the matrix elements
of the long-range interaction Hamiltonian~(\ref{HVDW}),
in the hyperfine-resolved basis, 
can also be accomplished with the 
help of the Wigner--Eckhart theorem,
as an alternative to the explicit construction
of states outlined in Sec.~\ref{sec2C}.
To this end, one writes the \vdw{} Hamiltonian,
given in Eq.~(\ref{HVDW}), as
\begin{eqnarray}
H_{\rm vdW} = 
- \frac{e^2}{4 \pi \epsilon_0} 
\frac{x_{A,-1} \, x_{B,+1} + x_{A,+1} \, x_{B,-1} +
2 x_{A,0} \, x_{B,0}}{R^3} ,
\end{eqnarray}
where the coordinates, in the spherical basis, are
\begin{eqnarray}
\label{xq}
x_{A,+1} = -\frac{1}{\sqrt{2}} \, (x_A + \ii \, y_A) \,,
\quad
x_{A,-1} = \frac{1}{\sqrt{2}} \, (x_A - \ii \, y_A)\, ,
\quad
x_{A,0} = z_A \,,
\end{eqnarray}
and same for atom $B$.

The unperturbed states are of the form
$| n, \ell, J, F, m_F, (S), (I) \rangle$
where we have previously defined 
the states as $| n, \ell, J, F, m_F \rangle$
with all quantum numbers being explained 
previously. The ``hidden'' quantum numbers 
are the electron spin $S$,
and the nuclear spin $I$. For hydrogen,
these attain the values $S= I = \frac12$
and are the same for all hydrogen states 
being discussed here. Still, the 
quantum numbers $S$ and $I$  need to be 
taken into account in the vector recoupling which will
be described in the following.
First, one eliminates the magnetic quantum numbers
$m_F$ and $m'_F$ by the Wigner--Eckhart theorem,
\begin{eqnarray}
\fl \langle n', \ell', J', F', m'_F, (S), (I) | \, T^1_q \, 
| n, \ell, J, F, m_F, (S), (I) \rangle 
= (-1)^{F'-m_F'}\left(
\begin{array}{ccc} F' & 1 & F \\ -m_F' & q & m_F \end{array}
\right) \nonumber\\
\times 
\langle n', \ell', J', F', (S), (I) || \, \vec T(1) \, || 
n, \ell, J, F, (S), (I) \rangle \,,
\end{eqnarray}
where $T^1_{q=-1,0,1}$ are the elements 
of a tensor, the specialization to the 
case $k = 1$ of a tensor $T^k_q$ of rank $k$,
and $\langle n', \ell', J', F', (S), (I) || \, \vec T(1) \, || 
n, \ell, J, F, (S), (I) \rangle$ is the reduced matrix element.
The nuclear and electronic degrees of freedom can
be separated using a 6$j$ symbol 
(vector recoupling coefficient) as described in 
Refs.~\cite{Ed1957,BrSa1994},
\begin{eqnarray}
\fl \langle n', \ell', J', F', (S), (I) || \, {\vec T}(1) \, 
|| n, \ell, J, F, (S), (I) \rangle 
= (-1)^{J'+I+F+1} \, \sqrt{(2F+1)(2F'+1)}  \nonumber\\
\times\left\{
\begin{array}{ccc} J' & F' & I \\ F & J & 1 \end{array}\right\}
 \langle n', \ell', J', (S) || \, {\vec T}(1) \, || 
n, \ell, J, (S) \rangle\,.
\end{eqnarray}
Another vector recoupling coefficient is needed in order 
to separate the orbital angular momentum of the 
electron from the electron spin,
\begin{eqnarray}
\fl \langle n', \ell', J', (S) || \, {\vec T}(1) \, ||
n, \ell, J, (S) \rangle 
&= (-1)^{L'+S+J+1} \,
\sqrt{(2J+1)(2J'+1)}
\nonumber\\
&\times \left\{ 
\begin{array}{ccc} L' & J' & S \\ J & L & 1 \end{array}
\right\}
\langle n', \ell'|| \, {\vec T}(1) \, ||n, \ell \rangle\,.
\end{eqnarray}
The following results for the reduced matrix elements,
\numparts
\begin{eqnarray}
\langle n'= 1, \ell' = 0 || \, {\vec r} \, || n=4, \ell=1 \rangle =& \;
-3 \, \frac{ 2^{11}}{5^6} \, \sqrt{\frac35} \, a_0\,,
\\[0.1133ex]
\langle n'= 4, \ell' = 1 || \, {\vec r} \, || n=1, \ell=0 \rangle =& \;
3 \, \frac{ 2^{11}}{5^6} \, \sqrt{\frac35} \, a_0\,,
\\[0.1133ex]
\langle n'= 2, \ell' = 0 || \, {\vec r} \, || n=4, \ell=1 \rangle =& \;
- \frac{ 2^{9}}{3^6} \, \sqrt{\frac{10}{3}} \, a_0\,,
\\[0.1133ex]
\langle n'= 4, \ell' = 1 || \, {\vec r} \, || n=2, \ell=0 \rangle =& \;
\frac{ 2^{9}}{3^6} \, \sqrt{\frac{10}{3}} \, a_0\,,
\\[0.1133ex]
\langle n'= 2, \ell' = 0 || \, {\vec r} \, || n=2, \ell=1 \rangle =& \;
3 \, \sqrt{3} \, a_0\,,
\\[0.1133ex]
\langle n'= 2, \ell' = 1 || \, {\vec r} \, || n=2, \ell=0 \rangle =& \;
-3 \, \sqrt{3} \, a_0\,,
\end{eqnarray}
\endnumparts
for the rank one tensor $\vec r = \vec T(1)$,
cover all states relevant to the current investigation.
In order to evaluate the elements, one 
expresses them, after the application of the 
Wigner--Eckhart theorem, in terms of radial integrals
involving the standard hydrogenic bound-state 
wave functions~\cite{LaLi1958vol3,BeSa1957}.

%
% $4P$--$1S$ Interaction
%
\section{\texorpdfstring{$\bm{4P}$--$\bm{1S}$}{4P--1S} Interaction}
\label{sec3}
\begin{table*}
\begin{center}
%\begin{minipage}{0.9\linewidth}
%\captionsetup{width=0.9\linewidth}
\caption{\label{table1} Multiplicities in the $4P_{1/2}$--$4P_{3/2}$--$1S$ system.
One might wonder why $\frakF_z = \pm 3$ is possible for $F=2$. The 
answer is that $F=2$ here refers to the total angular momentum
(electron orbital plus electron spin plus nuclear spin)
of one of the atoms, while $\frakF_z = \pm 3$ refers to the 
angular momentum projection of the sum of the 
total angular momenta of both electrons i.e., $\frakF_z=F_{z,A}+F_{z,B}$.}
\begin{tabular}{ccccc}
%% \begin{tabular}{c@{\hspace{0.5cm}}c@{\hspace{0.5cm}}c@{\hspace{0.5cm}}c@{\hspace{0.5cm}}c}
\hline
\hline
\stdrule
                & $\frakF_z=0$   &   $\frakF_z=\pm 1$ &    $\frakF_z=\pm 2$ &   $\frakF_z=\pm 3$ \\
\hline
\hline
\stdrule
$(J=\frac32,F=2)$   &    8       &      8    &       6    &      2     \\
\stdrule
$(J=\frac32,F=1)$   &    8       &      6    &       2    &      0     \\
\hline
\stdrule
$(J=\frac32)$       &    16      &     14    &       8    &     2    \\
\hline
\stdrule
$(J=\frac12,F=1)$   &     8      &      6    &       2    &      0     \\
\stdrule
$(J=\frac12,F=0)$   &     4      &      2    &       0    &      0     \\
\hline
\stdrule
$(J=1/2)$       &    12      &      8    &       2    &      0     \\
\hline
\hline
\stdrule
$(J=\frac12)+(J=\frac32)$  & 28      &     22    &      10    &      2     \\
\hline
\hline
\end{tabular}
%\end{minipage}
\end{center}
\end{table*}

%
% Selection of the States
%
\subsection{Selection of the States}
\label{sec3A}

The task is to diagonalize the Hamiltonian
given in Eq.~(\ref{H}),
\begin{eqnarray}
H = H_{\LS,A} + H_{\LS,B} +
H_{\HFS,A} + H_{\HFS,B} +
H_{\fs,A} + H_{\fs,B} +
H_{\vdW} \,,
\end{eqnarray}
in a quasi-degenerate basis, for two 
atoms, the first being in a $4P$ state,
the second being in a substate of the $1S$ hyperfine manifold.
Retardation does not need to be considered.
According to Refs.~\cite{HoHo2016,hdel},
the $4P$ fine-structure frequency
$\nu_{\rm FS} = \nu(4P_{3/2} - 4P_{1/2})$, 
\begin{eqnarray}
\nu_{\rm FS} \approx 1\,371 \, {\rm MHz} \,,
\end{eqnarray}
approximately coincides with the 
$1S$ hyperfine-structure frequency
\begin{eqnarray}
\nu_{\rm HFS} \approx 1\,420 \, {\rm MHz} \,,
\end{eqnarray}
which is the $21 \, {\rm cm}$ line.
Hence, in order to be self-consistent, we need to 
include both the $4P_{1/2}$ as well as the
$4P_{3/2}$ states into our hyperfine-resolved 
basis.

We select the $(4P)_A (1S)_B$ and 
$(1S)_A (4P)_B$ states, with all hyperfine levels
resolved, from the respective manifolds, and
obtain the following total multiplicities 
when all $4P_{1/2}$ and $4P_{3/2}$ states
are added into the basis (see also Table~\ref{table1})
\begin{eqnarray}
\eqalign{
g(\frakF_z = \pm 3) = \; 2  \,, \qquad
g(\frakF_z = \pm 2) = \; 10 \,, \cr
g(\frakF_z = \pm 1) = \; 22 \,, \qquad
g(\frakF_z = 0) = \; 28  \,. }
\end{eqnarray}
The multiplicities are the sums of the multiplicities in the 
$4P_{3/2}$--$1S$ system,
\begin{eqnarray}
\eqalign{
g(J = \frac32, \frakF_z = \pm 3) = \; 2 \,, \qquad
g(J = \frac32, \frakF_z = \pm 2) = \; 8 \,, \cr
g(J = \frac32, \frakF_z = \pm 1) = \; 14 \,, \qquad
g(J = \frac32, \frakF_z = 0) = \; 16 \,, }
\end{eqnarray}
and in the $4P_{1/2}$--$1S$ system,
\begin{eqnarray}
\fl
g(J = \frac12, \frakF_z = \pm 2) = \; 2 \,, \quad
g(J = \frac12, \frakF_z = \pm 1) = \; 8 \,, \quad
g(J = \frac12, \frakF_z = 0) = \; 12 \,.
\end{eqnarray}
We work with the full $J = \frac12$ and $J = \frac32$
manifolds throughout our investigation.

%
% Matrix Elements
%
\subsection{Matrix Elements of the Total Hamiltonian}
\label{sec3B}

Matrix elements of the total 
Hamiltonian~(\ref{H}) now have to be computed 
in the space spanned by the two-atom states,
which are given in 
Eqs.~(\ref{4p-singlet})--(\ref{4Pstates}) (for the $4P_{1/2}$ states),
as well as the $4P_{3/2}$ states.
These elements may either be determined 
by a computer symbolic program~\cite{Wo1999}
or using the Wigner-Eckart procedure described in Sec.~\ref{sec2D}.
It is useful to define the parameters
\numparts
\begin{eqnarray}
\label{parameters}
\calH &\equiv 
\frac{\alpha^4}{18} \, g_p \, \frac{m_e }{m_p}\,m_e \,c^2 
= h \, 59.21498 \, {\rm MHz}
\,,\label{eq:HFSplit}\\
\calL_2 &\equiv h \, 1057.845(9) \, {\rm MHz}
\,, \label{eq:LS2Split}\\
%
%\color{red}
\calL_4 &\equiv 
%\color{red}
h \times \frac{1}{8}\, \times1057.845(9)\, {\rm MHz}
\,, \label{eq:LS4Split}\\
%\color{black}
%
\calF &\equiv \frac{\alpha^4 \, m_e  \, c^2}{256} 
= h \, 1368.660 \, {\rm MHz}
\,, \label{eq:LSSplit}\\
\calV (\rho) &\equiv 3\, \frac{e^2}{4 \pi \epsilon_0} \,
\frac{a_0^2}{R^3} = \frac{ 3 \, E_h }{ \rho^3 } \,,
\label{eq:InterV} 
\\
\calW(\rho) &\equiv \frac{3 \times 2^{22}}{5^{13}} \, 
\frac{e^2}{4 \pi \epsilon_0} \,
\frac{a_0^2}{R^3} = \frac{3 \times 2^{22}}{5^{13}} \, 
\frac{ E_h }{ \rho^3 } \,,
\label{eq:InterW}
\end{eqnarray}
\endnumparts
where $R = a_0 \rho$, and $a_0$ is the Bohr radius,
$\calL_4$ is the $4S-4P_{1/2}$ Lamb shift, and
$E_h=\alpha^2 m_e c^2$ is the Hartree energy.
Our symbol $\calH$ is equivalent 
to one-third of the hyperfine splitting of 
the $2S$ state~\cite{KoEtAl2009},
while $\calL_2$ is the $2S$--$2P_{1/2}$ Lamb shift~\cite{LuPi1981}.
The interaction energy 
$\calV (\rho)$ depends on the interatomic separation $R$,
viz., $\rho$. 
We have used the identity
\begin{eqnarray}
\fl 
\frac{e^2}{4 \pi \epsilon_0} \, \frac{a_0^2}{R^3}
&= \;
\frac{4 \pi \alpha \epsilon_0 \hbar c}%
{4 \pi \epsilon_0} \, \frac{\alpha m_e c}{\hbar} \, 
\frac{1}{\rho^3}
= \; \frac{\alpha \hbar c \, \alpha m_e c}{\hbar} \, \frac{1}{\rho^3}
=\frac{\alpha^2 m_e c^2}{\rho^3} = \frac{E_h}{\rho^3} \,.
\end{eqnarray}

The natural scale for the constants $\calH$ and $\calL$ is 
an energy of order $\alpha^3 \, E_h$. Hence, we write
\begin{eqnarray}
\calH = \; \alpha^3 \, E_h \, C_{\calH} \,,
\qquad
\calL_n = \; \alpha^3 \, E_h \, C_{\calL,n} \,,\qquad
\calF =& \; \alpha^3 \, E_h \, C_{\calF} \,,
\end{eqnarray}
where we set $C_{\calH}=g_p/18\alpha  \times (m_e/m_p) = 0.0231596$,  
$C_{\calF}=1/256\alpha = 0.5352969$ and  $C_{\calL,4}=C_{\calL,2}/8=0.0517167 $.
Then, we can write for typical expressions of
second-order energy shifts,
\begin{eqnarray}
\frac{\calV^2(\rho)}{T_1 \calH + T_2 \calL_n +T_3\calF} 
=\frac{9}{T_1 \, C_{\calH} + T_2 \,C_{\calL,n}+T_3C_{\calF}} \, \,
\frac{E_h}{\alpha^3 \,\rho^6} \,,
\end{eqnarray}
where $T_1$, $T_2$  and $T_3$ typically are rational fractions,
to be determined by separate calculations.

A particularly interesting feature is that the hyperfine 
Hamiltonian actually is not diagonal in the space of the 
$4P_{1/2}$ and $4P_{3/2}$ states.
Rather, one has a mixing among the $F=1$ states of the
$4 P_{1/2}$ and $4 P_{3/2}$ manifolds,
with the mixing matrix element being given by
(see Ref.~\cite{Pa1996mu} for an outline of the 
calculation)
\begin{eqnarray}
\langle 4P_{3/2}^{F=1} (F_z) | H_{\rm HFS} |
4P_{1/2}^{F=1} (F_z)  \rangle = X \,.
\end{eqnarray}
We restrict the discussion here to one atom only,
say, atom $A$, omitting the subscript on 
$H_{\rm HFS} \equiv H_{{\rm HFS},A}$.
For the two states to 
be coupled, the magnetic projection $F_z$ has to be the same, though.
Otherwise, the matrix element vanishes.
Thus, in the basis of states
\begin{eqnarray}\fl
| a \rangle = \; | 4P_{1/2}^{F=1} F_z = 1) \rangle 
= | 4, 1, \frac12, 1, 1 \rangle \,, \quad
| b \rangle = \; | 4P_{1/2}^{F=1} (F_z = 0) \rangle 
= | 4, 1, \frac12, 1, 0 \rangle \,, \\
\fl
| c \rangle = \; | 4P_{1/2}^{F=1} F_z = -1) \rangle 
= | 4, 1, \frac12, 1, -1 \rangle \,, \quad
| d \rangle = \; | 4P_{3/2}^{F=1} (F_z = 1) \rangle 
= | 4, 1, \frac32, 1, 1 \rangle \,, \nonumber\\
\fl
| e \rangle = \; | 4P_{3/2}^{F=1} (F_z = 0) \rangle 
= | 4, 1, \frac32, 1, 0 \rangle \,, \quad
| f \rangle = \; | 4P_{3/2}^{F=1} (F_z = -1) \rangle
= | 4, 1, \frac32, 1, -1 \rangle \,,\nonumber
\end{eqnarray}
the matrix of the Hamiltonian $H_{\rm HFS} + H_{\rm FS}$
is evaluated as
\begin{eqnarray}
\label{HFSFSmix}
H_{\rm HFS+FS} = 
\left( \begin{array}{cccccc}
D & 0 & 0 & X & 0 & 0 \\
0 & D & 0 & 0 & X & 0 \\
0 & 0 & D & 0 & 0 & X \\
X & 0 & 0 & -D+\calF & 0 & 0 \\
0 & X & 0 & 0 & -D+\calF & 0 \\
0 & 0 & X & 0 & 0 & -D+\calF \\
\end{array} \right),
\end{eqnarray}
where 
\begin{eqnarray}
D = g_p \frac{\alpha^4 m_e^2c^2}{576 \, m_p} \,,
\qquad
X = -g_p \frac{\alpha^4 m_e^2  c^2}{1152 \, \sqrt{2} \, m_p} \,.
\end{eqnarray}
Here, $g_p$ is the proton $g$ factor,
while $D$ is a diagonal matrix element, and $X$ is the off-diagonal
element given above.

The $6\times 6$ Hamiltonian matrix~(\ref{HFSFSmix})
can be decomposed into three identical submatrices corresponding
to $F_z = -1, 0$ and $+1$. Each submatrix is of dimension two,
e.g., the one spanned by $|a\rangle$ and $|d\rangle$.
The Hamiltonian matrix is
\begin{eqnarray}
H^{F_z =1}_{\rm HFS+FS}=\left(
\begin{array}{cc}
 D & X \\
 X & -D+\calF \\
\end{array}
\right).
\end{eqnarray}
The eigenvalues of $H^{F_z = 1}_{\rm HFS+FS}$ are given by
\numparts
\begin{eqnarray}
\mathcal{E}_{+}=-D +\calF +\frac{X^2}{\calF-2D} +\Or(X^4)\,,\\
\mathcal{E}_{-}=D-\frac{X^2}{\calF-2D} +\Or(X^4).
\end{eqnarray}
\endnumparts
The second-order shift in the eigenvalues, 
$\Delta = X^2/(\calF-2D)$, is numerically 
equal to $ \num{4.7659e-14} E_h$, 
where $E_h= \alpha^2 m_e c^2$ is the Hartree energy.
For simplicity, we thus define the parameter
\begin{eqnarray} \label{DefineDelta}
\Delta = \num{4.7659e-14} \,,
\qquad
\frac{\Delta \cdot E_h}{h} = \SI{313.58}{\hertz} \,.
\end{eqnarray}
The normalized eigenvectors of $H^{F_z = 1}_{\rm HFS-FS}$ are
\numparts
\begin{eqnarray}
|\varphi_{+}\rangle =
\frac{1}{\sqrt{\alpha_{-}^2+1}}\left(\alpha_{-}\,|a\rangle +|d\rangle\right)\,,\\
|\varphi_{-}\rangle =
\frac{1}{\sqrt{\alpha_{+}^2+1}}\left(\alpha_{+}\,|a\rangle +|d\rangle\right)\,,
\end{eqnarray}
\endnumparts
where the coefficients $\alpha_{\pm}$ are given by
\begin{eqnarray}
\alpha_{\pm}=\frac{2 D-\calF \pm
\sqrt{4 (D^2- D \calF+ X^2) + \calF^2}}{2 X}\,.
\end{eqnarray}

%% Note that calV is called \[CapitalDelta]dd in the notebooks.
Examples of expectation values
of the hyperfine $H_\HFS$ and Lamb shift $H_\LS$ Hamiltonians (for states of
both atoms $A$ and $B$) are 
%
%\allowdisplaybreaks
\numparts
\begin{eqnarray} F_z
\langle n, \ell, J, F, F_z | H_\LS| n, \ell, J, F, F_z\rangle = \; 
\mathcal{L}_n \, \delta_{\ell 0},
\\
\langle 1, 0, \frac12, 1,F_z| H_\HFS | 1, 0, \frac12, 1,F_z \rangle = \; 
6 \, \mathcal{H} \,,
\\
\langle 1, 0, \frac12, 0, 0 | H_\HFS | 1, 0, \frac12, 0, 0 \rangle = \; 
-18 \, \mathcal{H} \,,
\\
\langle 4, 1, \frac12, 1,F_z | H_\HFS | 4, 1, \frac12, 1, F_z\rangle = \; 
\frac{1}{32} \, \mathcal{H} \,,
\\
\langle 4, 1, \frac12, 0, 0 | H_\HFS | 4, 1, \frac12, 0, 0 \rangle = \; 
-\frac{3}{32} \, \mathcal{H} \,.
\\
\langle 4, 1, \frac32, 2, F_z | H_\HFS | 4, 1, \frac32, 2, F_z \rangle = \;
\frac{3}{160} \, \mathcal{H} \,,
\\
\langle 4, 1, \frac32, 1,F_z| H_\HFS | 4, 1, \frac32, 1,F_z\rangle = \;
-\frac{1}{32} \, \mathcal{H} \,.
\end{eqnarray}
\endnumparts
The hyperfine splitting energy between $4P_{1/2}(F=1)$ and 
$4P_{1/2}(F=0)$ states thus amounts to $\mathcal{H}/8$, 
while between $4P_{3/2}(F=2)$ and $4P_{3/2}(F=1)$ states,
it is $\mathcal{H}/20$.
The $1S$-state hyperfine splitting is $24 \mathcal{H}$.
For the product state of atoms $A$ and $B$,
we shall use the notation
\begin{eqnarray}
| (n_A, \ell_A, J_A, F_A, F_{z,A})_A \, 
(n_B, \ell_B, J_B, F_B, F_{z,B})_B \, \rangle \,,
\end{eqnarray}
which summarizes the quantum numbers of both atoms.

\begin{figure*}[t!]
\begin{center}
\begin{minipage}{0.91\linewidth}
\begin{center}
\includegraphics[width=0.9\textwidth]{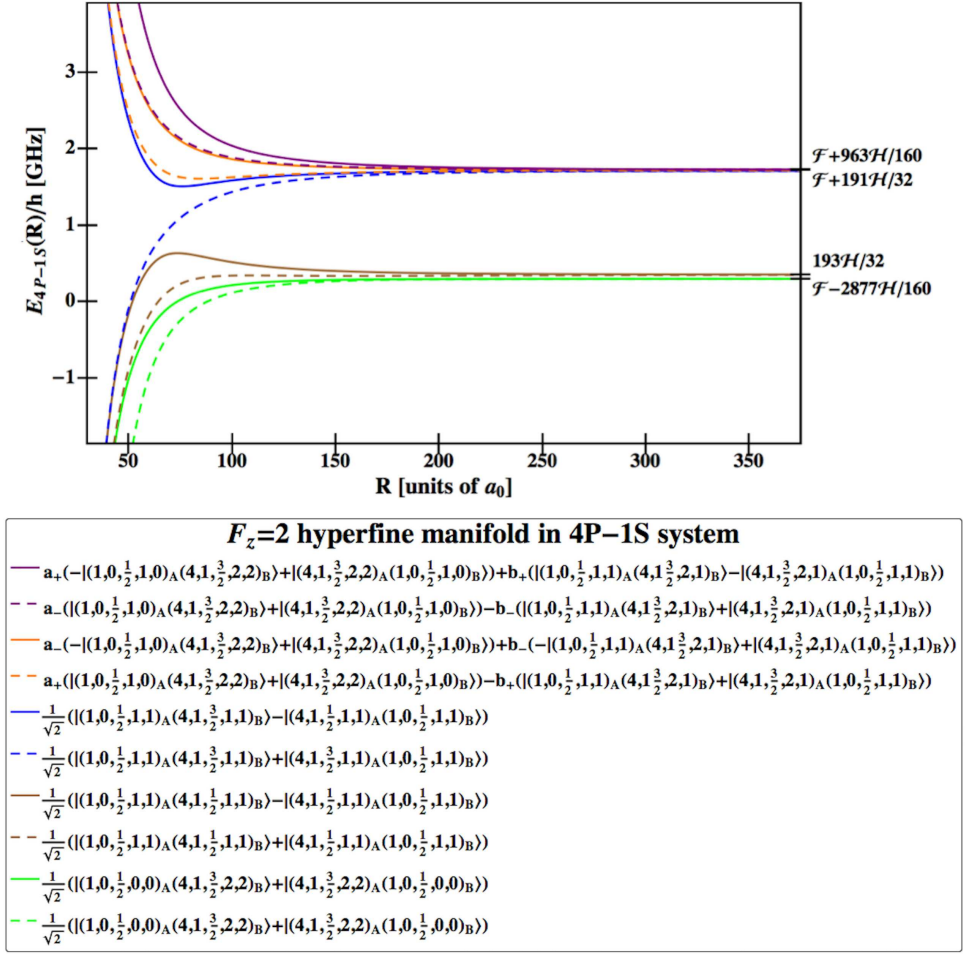}
%
%\captionsetup{width=0.9\linewidth}
\caption{\label{fig1}Evolution of the energy levels  as a
function of interatomic distance.  The  vertical axis is the energy 
divided by the Planck constant and given in units of 
$10^9 \, {\rm Hz}$ (GHz).  The interatomic 
separation in the horizontal  axis is in units of Bohr's radius. At large 
separation, there are  four energy levels, which match the  number of  
unperturbed energy  values of  matrix $H_{F_z = 2}$. As the interatomic 
distance decreases, the energy levels repel each other and 
are visually discernible.
The coefficients $a_{\pm}$ and $b_{\pm}$ are given 
by Eq.~(\ref{constant:ab}).}
\end{center}
\end{minipage}
\end{center}
\end{figure*}

%
% Manifold $F_z = 3$
%
\subsection{Manifold \texorpdfstring{${\frakF_z = 3}$}{Fz = 3}}
\label{sec3C}

States can be classified according to the 
quantum number $\frakF_z=F_{z,A}+F_{z,B}$, because the $z$ component of the 
total angular momentum 
commutes \cite{JeEtAl2017vdWii} with the total Hamiltonian
given in Eq.~(\ref{H}).
Within the $4P_{1/2}$--$4P_{3/2}$--${1S}_{1/2}$ system,
the states in the manifold $\frakF_z = 3$ are given as follows,
\begin{eqnarray}\fl
| \phi_{1} \rangle = \; | (1,0,\frac12, 1,1)_A \, (4,1, \frac32, 2,2)_B \rangle \,,
\quad
| \phi_{2} \rangle =& \; | (4,1,\frac32,2,2)_A \, (1,0,\frac12,1,1)_B \rangle.
\end{eqnarray}
In full analogy to the $1S$--$6P$ system
analyzed in Ref.~\cite{JeAd2017atoms},
we have ordered the basis vectors in ascending order of
the quantum numbers, starting from the last member in 
the list. The Hamiltonian matrix evaluates to 
\begin{eqnarray}
\label{matFz2}
H_{\frakF_z = 3} = 
\left(
\begin{array}{cc}
\frac{963}{160} \,\DHF{} + \calF & \frac{3 \times 2^{22}}{5^{13}} \DDD{}(\rho) \\[2ex]
\frac{3 \times 2^{22}}{5^{13}} \DDD{}(\rho) & \frac{963}{160} \,\DHF{} + \calF \\
\end{array}
\right) \,.
\end{eqnarray}
We have subtracted the sum of the Dirac energies of the 
$1S$ and $4P_{1/2}$ hyperfine centroids, and the $1S$ Lamb shift 
is absorbed in the definition of the $1S$ hyperfine centroid
energy, as outlined in Sec.~\ref{sec2B}.

The eigenenergies corresponding to $H_{\frakF_z = 3}$ 
are given as follows,
\begin{eqnarray}
\label{eigenenergy}
E_{\pm} (\rho)=
\frac{963}{160} \DHF{} + \calF\mp \frac{3 \times 2^{22}}{5^{13}} \DDD{}(\rho) \,,
\end{eqnarray}
with the corresponding eigenvectors,
\begin{eqnarray}
| u_\pm \rangle = \frac{1}{\sqrt{2}} \, 
( | \phi_{1} \rangle \pm | \phi_{2} \rangle ) \,.
\end{eqnarray}
The average of the first-order 
shifts (linear in $\DDD{}(\rho)$) vanishes.
The addition of the first-order shifts 
leads to exact energy eigenvalues [see Eq.~(\ref{eigenenergy})],
and it is thus not meaningful to analyze a 
potential second-order shift within the 
$\frakF_z = 3$ manifold.

%
% Manifold $\maybebm{\frakF_z = 2}$
%
\subsection{Manifold \texorpdfstring{${\frakF_z = 2}$}{Fz = 2}}
\label{sec3D}

We order the $10$ states in this manifold in 
order of ascending quantum numbers, 
%
%\allowdisplaybreaks
\numparts
\begin{eqnarray}
\label{XXXa}
\fl | \psi_{1} \rangle = \; | (1,0,\frac12,0,0)_A \, (4,1,\frac32,2,2)_B \rangle
\,,\quad
| \psi_{2} \rangle = \; | (1,0,\frac12,1,0)_A \, (4,1,\frac32,2,2)_B \rangle
\,, \\[0.1133ex]
\fl | \psi_{3} \rangle = \; | (1,0,\frac12,1,1)_A \, (4,1,\frac12,1,1)_B \rangle
\,,  \quad
| \psi_{4} \rangle = \; | (1,0,\frac12,1,1)_A \, (4,1,\frac32,1,1)_B \rangle
\,, \\[0.1133ex]
\fl | \psi_{5} \rangle = \; | (1,0,\frac12,1,1)_A \, (4,1,\frac32,2,1)_B \rangle
\,,  \quad
| \psi_{6} \rangle = \; | (4,1,\frac12,1,1)_A \, (1,0,\frac12,1,1)_B \rangle
\,, \\[0.1133ex]
\fl | \psi_{7} \rangle = \; | (4,1,\frac32,1,1)_A \, (1,0,\frac12,1,1)_B \rangle
\,,  \quad
| \psi_{8} \rangle = \; | (4,1,\frac32,2,1)_A \, (1,0,\frac12,1,1)_B \rangle
\,, \\[0.1133ex]
\label{XXXe}
\fl | \psi_{9} \rangle = \; | (4,1,\frac32,2,2)_A \, (1,0,\frac12,0,0)_B \rangle
\,, \quad
| \psi_{10} \rangle = \; | (4,1,\frac32,2,2)_A \, (1,0,\frac12,1,0)_B \rangle \,.
\end{eqnarray}
\endnumparts
States $| \psi_{3} \rangle$ and $| \psi_{6} \rangle$ are $4P_{1/2}$ states,
the rest are $4P_{3/2}$ states (see also the multiplicities indicated in 
Table~\ref{table1}). Among the $4P_{3/2}$ states, 
$| \psi_{4} \rangle$ and $| \psi_{7} \rangle$ have $F = 1$, the rest have 
$F = 2$. The Hamiltonian matrix is $10 \times 10$ and has the structure
\begin{eqnarray}
\label{matFz1}
H_{\frakF_z = 2} = \left(
\begin{array}{cc}
H_{AA} & H_{AB} \\
H_{AB}^{\rm T} & H_{BB} 
\end{array} \right) \,,
\end{eqnarray}
where $H_{AA}$, $H_{AB}$, and $H_{BB}$
are $5 \times 5$ matrices, of the form
\begin{eqnarray}
H_{AA} = \left(
\begin{array}{ccccc}
\FS4{}-\frac{2877 \DHF{}}{160} & 0 & 0 & 0 & 0  \\
0 & \frac{963 \DHF{}}{160}+\FS4{} & 0 & 0 & 0 \\
0 & 0 & \frac{193 \DHF{}}{32} & -\frac{\DHF{}}{64 \sqrt{2}} & 0 \\
0 & 0 & -\frac{\DHF{}}{64 \sqrt{2}} & \frac{191 \DHF{}}{32}+\FS4{} & 0 \\
0 & 0 & 0 & 0 & \frac{963 \DHF{}}{160} \\
\end{array}
\right) \,,
\end{eqnarray}
as well as
\begin{eqnarray}
\fl\qquad
H_{AB} = \left(
\begin{array}{ccccc}
-\sqrt{3} \WWW{} (\rho)& \sqrt{6} \WWW{} (\rho)& 0 & 0 & 0 \\
\sqrt{3} \WWW{} (\rho)& \sqrt{\frac{3}{2}} \WWW{} (\rho)& \frac{3}{\sqrt{2}\,} \WWW{}(\rho) & 0 & 0 \\
-2 \WWW{} (\rho)& -\sqrt{2} \WWW{} (\rho)& \sqrt{6} \WWW{} (\rho)& -\sqrt{3} \WWW{} (\rho)& \sqrt{3} \WWW{} (\rho)\\
-\sqrt{2} \WWW{} (\rho)& -\WWW{} (\rho)& \sqrt{3} \WWW{} (\rho)& \sqrt{6} \WWW{} (\rho)& \sqrt{\frac{3}{2}} \WWW{} (\rho)\\
\sqrt{6} \WWW{} (\rho)& \sqrt{3} \WWW{} (\rho)& -3 \WWW{} (\rho)& 0 & \frac{3 }{\sqrt{2}\,}\WWW{}(\rho) \\
\end{array}
\right) \,,
\end{eqnarray}
and
\begin{eqnarray}
H_{BB} = \left(
\begin{array}{ccccc}
\frac{193 \DHF{}}{32} & -\frac{\DHF{}}{64 \sqrt{2}} & 0 & 0 & 0 \\
-\frac{\DHF{}}{64 \sqrt{2}} & \frac{191 \DHF{}}{32}+\FS4{} & 0 & 0 & 0 \\
0 & 0 & \frac{963 \DHF{}}{160}+\FS4{} & 0 & 0 \\
0 & 0 & 0 & \FS4{}-\frac{2877 \DHF{}}{160} & 0 \\
0 & 0 & 0 & 0 & \frac{963 \DHF{}}{160}+\FS4{} \\
\end{array}
\right) \,.
\end{eqnarray}
One can easily draw an adjacency graph as described in 
Ref.~\cite{JeEtAl2017vdWii,AdDeJe2017aphb}  and convince oneself that there is no hidden 
symmetry in the  Hamiltonian matrix $H_{\frakF_z = 2}$  which would otherwise 
decompose  into irreducible submatrices. The Hamiltonian matrix, $H_{\frakF_z = 2}$, has four 
degenerate subspaces. Within the 
sub-space of doubly-degenerate unperturbed energy  
$\mathcal{F}-2877 \mathcal{H}/160$, there is no off-diagonal coupling 
proportional to $\mathcal{W}(\rho)$ in the first order,
implying that the energy shift 
has  an $R^{-6}$ dependence.  The degenerate subspace  given by  
$| \psi_{3} \rangle$ and $| \psi_{6} \rangle$ has a Hamiltonian matrix
\begin{eqnarray}
\label{matFz1subA}
H_{\frakF_z = 2}^{(A)}= \left(
\begin{array}{cc}
 \frac{193 \mathcal{H}}{32} & -2 \mathcal{W}(\rho) \\
 -2 \mathcal{W}(\rho) & \frac{193 \mathcal{H}}{32} \\
\end{array}
\right)\,.
\end{eqnarray}
The eigenvalues are 
\begin{eqnarray}
\label{SubAEigVal}
E_{\pm}^{(A)}(\rho)= \frac{193 \mathcal{H}}{32}\mp2 \mathcal{W}(\rho)\,,
\end{eqnarray}
with corresponding normalized eigenvectors 
\begin{eqnarray}
\label{SubAEigVec}
|\psi_{\pm}^{(A)}\rangle= \frac{1}{\sqrt{2}}\left(| \psi_{3} \rangle \pm| \psi_{6} \rangle\right)\,.
\end{eqnarray}
A third degenerate subspace is given by $| \psi_{4} \rangle$ and $| \psi_{7} \rangle$.
The Hamiltonian matrix is 
\begin{eqnarray}
\label{matFz1subB}
H_{\frakF_z = 2}^{(B)}= \left(
\begin{array}{cc}
 \mathcal{F}+\frac{191 \mathcal{H}}{32} & -\mathcal{W}(\rho) \\
 -\mathcal{W}(\rho) & \mathcal{F}+\frac{191 \mathcal{H}}{32} \\
\end{array}
\right)\,.
\end{eqnarray}
The eigenvalues are 
\begin{eqnarray}
\label{SubBEigVal}
E_{\pm}^{(B)}(\rho)= \mathcal{F}+\frac{191 \mathcal{H}}{32} \mp \mathcal{W}(\rho)\,,
\end{eqnarray}
with corresponding normalized eigenvectors 
\begin{eqnarray}
|\psi_{\pm}^{(B)}\rangle= \frac{1}{\sqrt{2}}\left(| \psi_{4} \rangle \pm| \psi_{7} \rangle\right)\,.
\end{eqnarray}

\begin{table}[t]
%\begin{center}
%\begin{minipage}{0.8\linewidth}
%\begin{center}
%\captionsetup{width=0.9\linewidth}
\caption{\label{table2} Average second-order \vdw{} shifts for $4P_J$ hydrogen atoms 
interacting with ground-state atoms. Entries marked with a long hyphen
(--) indicate unphysical combinations of $F$ and $\frakF_z$ values.
We denote the scaled interatomic distance by $\rho = R/a_0$ 
and give all energy shifts in atomic units, i.e., 
in units of the Hartree energy $E_h = \alpha^2 m_e c^2$. 
Recall that ${\frakF}_z=F_{z,A}+F_{z,B}$ 
of the two atom system. The notation $\Delta$ is defined in Eq.~(\ref{DefineDelta}).}
%\begin{indented}\item[ ]
%
\begin{tabular}{c@{\hspace{0.5cm}}c@{\hspace{0.5cm}}c@{\hspace{0.5cm}}c@{\hspace{0.5cm}}c}
\hline
\hline
\stdrule
                & $\frakF_z=0$   &   $\frakF_z= \pm 1$ &    $\frakF_z=\pm 2$ &   $\frakF_z= \pm 3$ \\
\hline
\hline
\lrgrule
$(J=3/2,F=2)$   &  $\frac{\num{4.439e5}}{\rho^6}$ 
                & $\frac{\num{3.601e5}}{\rho^6}$ & $\frac{\num{3.416e5}}{\rho^6}$ &  0 \\
\lrgrule
$(J=3/2,F=1)$   &  $\Delta-\frac{\num{4.702e5}}{\rho^6}$ 
                &  $\Delta-\frac{\num{5.177e5}}{\rho^6}$ 
                &  $\Delta-\frac{\num{1.059e6}}{\rho^6}$ &  --- \\
\lrgrule
$(J=1/2,F=1)$   &  $-\Delta+\frac{\num{7.653e4}}{\rho^6}$ 
                &  $-\Delta+\frac{\num{1.970e5}}{\rho^6}$ 
                &  $-\Delta+\frac{\num{3.377e4}}{\rho^6}$ & --- \\
\lrgrule
$(J=1/2,F=0)$   & $-\frac{\num{1.005e5}}{\rho^6}$ 
                & $-\frac{\num{4.783e5}}{\rho^6}$ 
                & ---  &  --- \\
\hline
\hline
\end{tabular}
%\end{indented}
%\end{center}
%\end{minipage}
%\end{center}
\end{table}

We also have a four-fold degenerate subspace composed of 
$| \psi_{2} \rangle$, $| \psi_{5} \rangle$ , 
$| \psi_{8} \rangle$ and $| \psi_{10} \rangle$. 
The Hamiltonian matrix is
{\small
\begin{eqnarray}
\label{matFz1subC}
H_{\frakF_z = 2}^{(C)}= \left(
\begin{array}{cccc}
 \mathcal{F}+\frac{963 \mathcal{H}}{160} & 0 & \frac{3 \mathcal{W}(\rho)}{\sqrt{2}} & 0 \\
 0 & \mathcal{F}+\frac{963 \mathcal{H}}{160} & -3 \mathcal{W}(\rho) & \frac{3 \mathcal{W}(\rho)}{\sqrt{2}} \\
 \frac{3 \mathcal{W}(\rho)}{\sqrt{2}} & -3 \mathcal{W}(\rho) & \mathcal{F}+\frac{963 \mathcal{H}}{160} & 0 \\
 0 & \frac{3 \mathcal{W}(\rho)}{\sqrt{2}} & 0 & \mathcal{F}+\frac{963 \mathcal{H}}{160} \\
\end{array}
\right)\,.
\end{eqnarray}
}
The eigenvalues are 
\numparts
\begin{eqnarray}
\label{SubCEigVal}
E_{1}^{(C)}(\rho)=&\;\mathcal{F}+\frac{963}{160} \mathcal{H} 
- \frac{3}{2}  \left(\sqrt{3}+1\right) \mathcal{W}(\rho)\,,\\[0.1133ex]
E_{2}^{(C)}(\rho)=&\;\mathcal{F}+\frac{963}{160} \mathcal{H} 
- \frac{3}{2}  \left(\sqrt{3}-1\right) \mathcal{W}(\rho)\,,\\[0.1133ex]
E_{3}^{(C)}(\rho)=&\;\mathcal{F}+\frac{963}{160} \mathcal{H} 
+ \frac{3}{2}  \left(\sqrt{3}-1\right) \mathcal{W}(\rho)\,,\\[0.1133ex]
E_{4}^{(C)}(\rho)=&\;\mathcal{F}+\frac{963}{160} \mathcal{H} 
+ \frac{3}{2}  \left(\sqrt{3}+1\right) \mathcal{W}(\rho)\,,
\end{eqnarray}
\endnumparts
with corresponding normalized eigenvectors 
\numparts
\begin{eqnarray}
|\psi_{1}^{(C)}\rangle=& a_{+} \left(| \psi_{2} \rangle +| \psi_{10} \rangle\right)
-b_{+}\left(| \psi_{5} \rangle +| \psi_{8} \rangle\right)\,, \\[0.1133ex]
|\psi_{2}^{(C)}\rangle=& a_{-} \left(-| \psi_{2} \rangle +| \psi_{10} \rangle\right)
+b_{-}\left(-| \psi_{5} \rangle +| \psi_{8} \rangle\right)\,, \\[0.1133ex]
|\psi_{3}^{(C)}\rangle=& a_{-} \left(| \psi_{2} \rangle +| \psi_{10} \rangle\right)
-b_{-}\left(| \psi_{5} \rangle +| \psi_{8} \rangle\right)\,, \\[0.1133ex]
|\psi_{4}^{(C)}\rangle=& a_{+} \left(-| \psi_{2} \rangle +| \psi_{10} \rangle\right)
+b_{+}\left(| \psi_{5} \rangle -| \psi_{8} \rangle\right)\,, 
\end{eqnarray}
\endnumparts
where
\begin{eqnarray}
\label{constant:ab}
a_{\pm} =\frac{1}{\sqrt{2 \left(3\pm\sqrt{3}\right)}}\quad \mathrm{and}\quad
b_{\pm} =\frac{\sqrt{3}\pm1}{2\sqrt{ 3\pm\sqrt{3}}}\,.
\end{eqnarray}
In Fig.~\ref{fig1},
we plot the evolution of the energy eigenvalues within the $\frakF_z=2$ manifold
with respect to interatomic separation.

Of particular interest are second-order \vdw{} shifts,
which occur in the ($\frakF_z=2$) manifold. 
The first and most detailed approach to this 
calculation involves keeping $J$ and $F$ fixed,
and averaging only 
over the magnetic projections. We consider the
entries in the fourth column of Table~\ref{table2}.
First, we observe that there are no
$4P_{1/2}$ states with $F=0$ in the manifold $\frakF_z = 2$, because
of angular momentum selection rules (we have $\frakF_z = 2$ 
and hence $F \geq 2$ for all states in the manifold). 
The averaging over the magnetic projections
for given $J$ and $F$ (and $\frakF_z$, of course) fixed,
involves the calculation of the arithmetic mean of the 
second-order energy shifts, after selecting from the 
states given in Eqs.~(\ref{XXXa})---(\ref{XXXe}) those 
two-atom states where the $4P$ atom has the 
required quantum numbers. 

For example, the average for $J=3/2$, $F=2$, and of course,
$\frakF_z = 2$, is given as
\begin{eqnarray}
\label{avg1}
\fl \quad \left< E(4P_{3/2}, F=2,\frakF_z =2 \right> = 
\frac{1}{4} \left[ E^{(2)}(\psi_2) + E^{(2)}(\psi_8) + E^{(2)}(\psi_9) + E^{(2)}(\psi_{10}) \right] \,,
\end{eqnarray}
where the $E^{(2)}(\psi_i)$ are the second-order energy shifts
of the states $\psi_i$ given in  Eqs.~(\ref{XXXa})---(\ref{XXXe}).
For reference, we also indicate that
\begin{eqnarray}
\label{avg2}
\fl \quad \left< E(4P_{3/2}, F=1,\frakF_z =2 \right> = 
\frac{1}{2} \left[ E^{(2)}(\psi_4) + E^{(2)}(\psi_7) \right] \,,
\\[0.1133ex]
\fl \quad \left< E(4P_{1/2}, F=1,\frakF_z =2 \right> = 
\frac{1}{2} \left[ E^{(2)}(\psi_3) + E^{(2)}(\psi_6) \right] \,.
\end{eqnarray}
With respect to Table~\ref{table2}, we also
observe that the $\Delta$ term, which is the HFS--FS mixing 
term, only occurs for the $F=1$ states, and vanishes
for the $F=2$ states. 

It is then possible to calculate a 
weighted average over the possible values of $F$ within the 
($\frakF_z = 2$) manifold, by applying the multiplicities 
incurred within the reference manifold. So, for example,
on the basis of Eqs.~(\ref{avg1}) and~(\ref{avg2}), we have
\begin{eqnarray}
\fl \quad \left< E(4P_{3/2}, \frakF_z =2 ) \right>_F
=\frac{4 \left< E(4P_{3/2}, F=2,\frakF_z =2 \right> +
2 \left< E(4P_{3/2}, F=1,\frakF_z =2 \right>}{6} \,.
\end{eqnarray}

Specifically, one obtains
%%]
\numparts
\begin{eqnarray}
\left< E(4P_{3/2}, \frakF_z = 2) \right>_F =
\left( \frac14 \Delta - \frac{\num{8.442e3}}{\rho^6} \right) \, E_h \,,
\end{eqnarray}
while
\begin{eqnarray}
\left< E(4P_{1/2}, \frakF_z=2) \right>_F =
\left( -\Delta + \frac{\num{3.377e4}}{\rho^6} \right) E_h \,.
\end{eqnarray}
\endnumparts
The weighted average vanishes,
\begin{eqnarray}
2 \left< E(4P_{1/2}, \frakF_z = 2) \right>_F +
8 \left< E(4P_{3/2}, \frakF_z = 2) \right>_F  = 0 \,.
\end{eqnarray}

\begin{table*}
%\begin{center}
%\begin{minipage}{0.8\linewidth}
\begin{center}
%\captionsetup{width=0.9\linewidth}
\caption{\label{table3} 
Multiplicities in the $4P_{1/2}$--$4P_{3/2}$--($4S$;$2P_{1/2}$)--$2S$--$1S$ system.
The entries in the first seven rows refer to the 
$4P_{1/2}$--$4P_{3/2}$--$2S$ system, and are the same as those for 
the $4P_{1/2}$--$4P_{3/2}$--$1S$ system given in Table~\ref{table1}.
The eighth row gives the number of added $(4S,2P_{1/2})$ states which 
complete the basis of quasi-degenerate basis.
Finally, we end up with multiplicities of $40$, $30$, $12$ and $2$ for 
$\frakF_z = 0, \pm 1, \pm 2, \pm 3$, respectively (ninth row).}
\begin{tabular}{ccccc}
%% \begin{tabular}{c@{\hspace{0.5cm}}c@{\hspace{0.5cm}}c@{\hspace{0.5cm}}c@{\hspace{0.5cm}}c}
\hline
\hline
\stdrule
                & $\frakF_z=0$   &   $\frakF_z=\pm 1$ &    $\frakF_z=\pm 2$ &   $\frakF_z=\pm 3$ \\
\hline
\hline
\stdrule
$(J=\frac32,F=2)$   &    8       &      8    &       6    &      2     \\
\stdrule
$(J=\frac32,F=1)$   &    8       &      6    &       2    &      0     \\
\hline
\stdrule
$(J=\frac32)$       &    16      &     14    &       8    & 	2   \\
\hline
\stdrule
$(J=\frac12,F=1)$   &     8      &      6    &       2    &      0     \\
\stdrule
$(J=\frac12,F=0)$   &     4      &      2    &       0    &      0     \\
\hline
\stdrule
$(J=1/2)$            &    12      &      8    &       2    &      0     \\
\hline
\hline
\stdrule
$(J=\frac12)+(J=\frac32)$  & 28      &     22    &      10    &      2     \\
\hline
\hline
\stdrule
$(4S,2P_{1/2})$ States     &    12      &      8    &       2    &      0     \\
\hline
\hline
\stdrule
Total \#{} of States   & 40      &     30    &      12    &      2     \\
\hline
\hline
\end{tabular}
\end{center}
%\end{minipage}
%\end{center}
\end{table*}

%
% Manifold $\maybebm{\frakF_z = 1}$
%
%
\subsection{Manifold \texorpdfstring{${\frakF_z = 1}$}{Fz = 1}}
\label{sec3E}

We present the $22$ states in this manifold in
order of ascending quantum numbers,
%
%\allowdisplaybreaks
\numparts
\begin{eqnarray}
\label{statesFz1}
\fl | \Psi_{1} \rangle =\;| (1,0,\frac12,0,0)_A \, (4,1,\frac12,1,1)_B \rangle\,, \quad
| \Psi_{2} \rangle =\;| (1,0,\frac12,0,0)_A \, (4,1,\frac32,1,1)_B \rangle\,, \\
\fl | \Psi_{3} \rangle =\;| (1,0,\frac12,0,0)_A \, (4,1,\frac32,2,1)_B \rangle\,, \quad
| \Psi_{4} \rangle =\;| (1,0,\frac12,1,-1)_A \, (4,1,\frac32,2,2)_B \rangle\,, \\
\fl | \Psi_{5} \rangle =\;| (1,0,\frac12,1,0)_A \, (4,1,\frac12,1,1)_B \rangle\,,\quad
| \Psi_{6} \rangle =\;| (1,0,\frac12,1,0)_A \, (4,1,\frac32,1,1)_B \rangle\,, \\
\fl | \Psi_{7} \rangle =\;| (1,0,\frac12,1,0)_A \, (4,1,\frac32,2,1)_B \rangle\,, \quad
| \Psi_{8} \rangle =\;| (1,0,\frac12,1,1)_A \, (4,1,\frac12,0,0)_B \rangle\,, \\
\fl | \Psi_{9} \rangle =\;| (1,0,\frac12,1,1)_A \, (4,1,\frac12,1,0)_B \rangle\,, \quad
| \Psi_{10} \rangle =\;| (1,0,\frac12,1,1)_A \, (4,1,\frac32,1,0)_B \rangle\,, \\
\fl | \Psi_{11} \rangle =\;| (1,0,\frac12,1,1)_A \, (4,1,\frac32,2,0)_B \rangle\,,\quad
| \Psi_{12} \rangle =\;| (4,1,\frac12,0,0)_A \, (1,0,\frac12,1,1)_B \rangle\,, \\
\fl | \Psi_{13} \rangle =\;| (4,1,\frac12,1,0)_A \, (1,0,\frac12,1,1)_B \rangle\,, \quad
| \Psi_{14} \rangle =\;| (4,1,\frac12,1,1)_A \, (1,0,\frac12,0,0)_B \rangle\,, \\
\fl | \Psi_{15} \rangle =\;| (4,1,\frac12,1,1)_A \, (1,0,\frac12,1,0)_B \rangle\,, \quad
| \Psi_{16} \rangle =\;| (4,1,\frac32,1,0)_A \, (1,0,\frac12,1,1)_B \rangle\,, \\
\fl | \Psi_{17} \rangle =\;| (4,1,\frac32,1,1)_A \, (1,0,\frac12,0,0)_B \rangle\,, \quad
| \Psi_{18} \rangle =\;| (4,1,\frac32,1,1)_A \, (1,0,\frac12,1,0)_B \rangle\,, \\
\fl | \Psi_{19} \rangle =\;| (4,1,\frac32,2,0)_A \, (1,0,\frac12,1,1)_B \rangle\,,\quad
| \Psi_{20} \rangle =\;| (4,1,\frac32,2,1)_A \, (1,0,\frac12,0,0)_B \rangle\,, \\
\fl | \Psi_{21} \rangle =\;| (4,1,\frac32,2,1)_A \, (1,0,\frac12,1,0)_B \rangle\,, \quad
| \Psi_{22} \rangle =\;| (4,1,\frac32,2,2)_A \, (1,0,\frac12,1,-1)_B \rangle.
\end{eqnarray}
\endnumparts
We refer to Table~\ref{table2}
for the averaged second-order \vdw{} shifts in the $\frakF_z = 0$, 
$\frakF_z = +1$, $\frakF_z = +2$,  and $\frakF_z = +3$ manifolds.
The Hamiltonian matrix for  $\frakF_z = -3$ manifold is 
identical to that of $\frakF_z = +3$. The  $\frakF_z = -2$  manifold has  identical 
diagonal entries to that of $\frakF_z = +2$, while 
some off-diagonal entries are different. The same is true of the
$\frakF_z=\pm1$ manifolds. Yet,
the Born--Oppenheimer energy curves for $\frakF_z = \pm2$ 
and $\frakF_z=\pm 1$ are alike.

%\color{black}

%
% Second--Order Energy Shifts
%
%
\subsection{Second--Order Energy Shifts}
\label{sec3F}

As a function of $J$ and $F$, 
within the $4P$--$1S$ system, 
a global averaging over all possible $\frakF_z$ values for 
given $F$, leads to the results
\numparts
\begin{eqnarray}
\left< E(4P_{1/2}, F=0) \right>_{\frakF_z} &=& -\frac{\num{2.894e5}}{\rho^6} \, E_h \,,\\
\left< E(4P_{1/2}, F=1) \right>_{\frakF_z} &=& 
\left( -\Delta + \frac{\num{1.296e5}}{\rho^6} \right) E_h \,, \\
\left< E(4P_{3/2}, F=1) \right>_{\frakF_z} &=&
\left( \Delta - \frac{\num{5.920e5}}{\rho^6} \right) E_h \,, \\
\left< E(4P_{3/2}, F=2) \right>_{\frakF_z} &=& 
\frac{\num{3.353e5}}{\rho^6} \, E_h \,. 
\end{eqnarray}
\endnumparts
Comparing to Table~\ref{table2}, this average 
would correspond to an average over the entries in the 
different rows, for given~$F$.
A remark is in order. According to Eq.~(\ref{choice}),
we align the 
quantization axis with the straight line joining the two atoms;
this is the most natural choice. 
Of course, the precise identification of levels
with specific $\frakF_z$ components depends on the 
choice of the quantization axis. 
However, results for other orientations can be obtained
after the application of appropriate rotation matrices 
[see Chap.~2 of Ref.~\cite{BrSa1994} 
and Chap.~4 of Ref.~\cite{Ed1974ang}].
After averaging over the quantum numbers
$F_{z,A}$ and $F_{z,B}$, or, equivalently, 
the two-atom sum $\frakF_z$, the results are
independent of the choice of the quantization axis,
in view of the unitarity of the rotation matrices.

One can also average over the possible orientations of $F$,
namely, $F = J \pm \frac12$, for given $J$ and $\frakF_z$.
This amounts to an averaging over the first two entries in the
columns, and the third and fourth entry in every column,
of Table~\ref{table2}.
The results are
\numparts
\begin{eqnarray}
\left< E(4P_{1/2}, \frakF_z=0) \right>_F & = &
\left( -\frac23  \Delta + \frac{\num{1.752e4}}{\rho^6} \right) E_h \,, \\
\left< E(4P_{1/2}, \frakF_z=\pm 1) \right>_F & = &
\left( -\frac34 \Delta + \frac{\num{2.819e4}}{\rho^6} \right) E_h , \\
\left< E(4P_{1/2}, \frakF_z=\pm 2) \right>_F & = &
\left( -\Delta + \frac{\num{3.377e4}}{\rho^6} \right) E_h \,, 
\end{eqnarray}
\endnumparts
and
\numparts
\begin{eqnarray}
\left< E(4P_{3/2}, \frakF_z=0) \right>_F & =& 
\left( \frac12 \, \Delta - \frac{\num{1.131e4}}{\rho^6} \right) E_h \,, \\
\left< E(4P_{3/2}, \frakF_z=\pm 1) \right>_F & =& 
\left( \frac37 \, \Delta - \frac{\num{1.611e4}}{\rho^6} \right) E_h \,, \\
\left< E(4P_{3/2}, \frakF_z=\pm 2) \right>_F & =& 
\left( \frac14 \, \Delta - \frac{\num{8.442e3}}{\rho^6} \right) E_h \,, \\
\left< E(4P_{3/2}, \frakF_z=\pm 3) \right>_F & =& \; 0 \; \,.
\end{eqnarray}
\endnumparts
As a function of $J$, averaging over $F$ and $\frakF_z$ leads to the
results
\numparts
\begin{eqnarray}
\label{4P1S_4P12}
\left< E(4P_{1/2}) \right>_{F,\frakF_z} & =& 
\left( -\frac34 \, \Delta + \frac{\num{2.489e4}}{\rho^6} \right) E_h \,, \\
\label{4P1S_4P32}
\left< E(4P_{3/2}) \right>_{F, \frakF_z} &=& 
\left( \frac38 \, \Delta - \;  \frac{\num{1.245e4}}{\rho^6} \right) E_h \,.
\end{eqnarray}
\endnumparts
Without hyperfine resolution, there are four $J = \frac32$ states
and two $J = \frac12$ states. Hence, the fine-structure 
average of the latter two results vanishes.

%
% $2S$--$4P_{1/2}$ Interaction
%
\section{\texorpdfstring{$\bm{4P}$--$\bm{2S}$}{4P--2S} Interaction}
\label{sec4}

%
% Selection of the States
%
\subsection{Selection of the States}

The analysis of the interaction of 
excited $4P$ hydrogen atoms with metastable $2S$ atoms
is more complicated than that with 
ground-state atoms.
The reason is that we cannot simply restrict the basis of states to 
the $4P_{1/2}$, $4P_{3/2}$, and $2S$ states,
and just replace the $1S$ state from the 
previous calculation with the metastable $2S$
states. One observes that $| (4P)_A (2S)_B \rangle$ states are energetically 
quasi-degenerate with respect to $| (4S)_A (2P_{1/2})_B \rangle$
states, and removed from each other only by the 
classic $2S$--$2P_{1/2}$ Lamb shift.
It is thus necessary to augment the basis of 
states by the $4S$--$2P_{1/2}$ states,
and to carry out a full analysis for 
the $4P_{1/2}$--$4P_{3/2}$--($4S$;$2P_{1/2}$)--$2S$ system.
The notation indicates that 
the $4S$--$2P_{1/2}$ states are merely added as 
virtual states, for the calculation of second-order
energy shifts. 

Due to selection rules, we may reduce 
the number of states in the basis, 
according to Table~\ref{table3}.
Because the total Hamiltonian~(\ref{H}) 
commutes with the total angular momentum $\vec F$,
we obtain multiplicities of 
$28$, $22$, $10$ and $2$, for the 
manifolds with $\frakF_z = 0$, $\frakF_z = \pm 1$, $\frakF_z = \pm 2$, 
and $\frakF_z = \pm 3$.
However, the addition of the $(4S; 2P_{1/2})$ states
finally leads to multiplicities of
$40$, $30$, $12$ and $2$, for the 
manifolds with $\frakF_z = 0$, $\frakF_z = \pm 1$, $\frakF_z = \pm 2$, 
and $\frakF_z = \pm 3$.

\begin{table*}
\begin{center}
%\begin{minipage}{0.8\linewidth}
\begin{center}
%\captionsetup{width=0.9\linewidth}
\caption{\label{table4} Average second-order \vdw{} shifts for $4P_J$ hydrogen atoms
interacting with $2S$ metastable atoms. Entries marked with a long hyphen
(---) indicate unphysical combinations of $F$ and $\frakF_z$ values.
We denote the scaled interatomic distance by $\rho = R/a_0$
and give all energy shifts in atomic units, i.e.,
in units of the Hartree energy $E_h = \alpha^2 m_e c^2$. 
The notation $\Delta$ is defined in Eq.~(\ref{DefineDelta}).}
\begin{tabular}{c@{\hspace{0.5cm}}c@{\hspace{0.5cm}}c@{\hspace{0.5cm}}c@{\hspace{0.5cm}}c}
\hline
\hline
\stdrule
                & $\frakF_z=0$   &   $\frakF_z= \pm 1$ &    $\frakF_z=\pm 2$ &   $\frakF_z= \pm 3$ \\
\hline
\hline
\lrgrule
$(J=3/2,F=2)$   &  $\frac{\num{4.800e9}}{\rho^6}$
                &  $\frac{\num{3.996e9}}{\rho^6}$ 
                &  $\frac{\num{2.194e9}}{\rho^6}$ &  0 \\
\lrgrule
$(J=3/2,F=1)$   &  $\Delta+\frac{\num{3.973e9}}{\rho^6}$
                &  $\Delta+\frac{\num{2.947e9}}{\rho^6}$
                &  $\Delta+\frac{\num{6.966e8}}{\rho^6}$ &  --- \\
\lrgrule
$(J=1/2,F=1)$   &  $-\Delta+\frac{\num{8.916e9}}{\rho^6}$
                &  $-\Delta+\frac{\num{7.904e9}}{\rho^6}$
                &  $-\Delta+\frac{\num{4.976e9}}{\rho^6}$ & --- \\
\lrgrule
$(J=1/2,F=0)$   & $\frac{\num{8.216e9}}{\rho^6}$
                & $\frac{\num{7.302e9}}{\rho^6}$
                & ---  &  --- \\
\hline
\hline
\end{tabular}
\end{center}
%\end{minipage}
\end{center}
\end{table*}

%
% Second--Order Energy Shifts
%
\subsection{Second--Order Energy Shifts}
\label{sec4B}

In Table~\ref{table4}, we present results for 
second-order energy shifts 
within the individual $(J, F, \frakF_z)$ manifolds.
For individual $J$ and $F$ quantum 
numbers, an averaging over the magnetic quantum
projections $\frakF_z$ leads to the results
\numparts
\begin{eqnarray}
\left< E(4P_{1/2}, F=0) \right>_{\frakF_z} &=& \frac{\num{7.759e9}}{\rho^6} \, E_h \,, \\
\left< E(4P_{1/2}, F=1) \right>_{\frakF_z} &=& 
\left( \Delta + \frac{\num{7.753e9}}{\rho^6} \right) E_h \,, \\
\left< E(4P_{3/2}, F=1) \right>_{\frakF_z} &=& 
\left( -\Delta + \frac{\num{2.914e9}}{\rho^6} \right) E_h \,, \\
\left< E(4P_{3/2}, F=2) \right>_{\frakF_z} &=& \frac{\num{3.216e9}}{\rho^6} \, E_h \,.
\end{eqnarray}
\endnumparts
These results can be obtained from the entries in Table~\ref{table4},
weighing the terms with the multiplicities given in Table~\ref{table3}
(for an averaging over the rows).

Alternatively, one 
may opt to average over the possible orientations of $F$,
namely, $F = J \pm \frac12$, for given $J$ and $\frakF_z$.
This procedure is equivalent to an averaging over the 
first two entries in the 
columns (two possible orientations for $F$), 
and the third and fourth entry in every column,
of Table~\ref{table4}. The results then read as 
\numparts
\begin{eqnarray}
\left< E(4P_{1/2}, \frakF_z=0) \right>_F &=&
\left( -\frac23  \Delta + \frac{\num{8.682e9}}{\rho^6} \right) E_h \,, \\
\left< E(4P_{1/2}, \frakF_z=\pm 1) \right>_F &=& 
\left( -\frac34  \Delta + \frac{\num{7.754e9}}{\rho^6} \right) E_h , \\
\left< E(4P_{1/2}, \frakF_z=\pm 2) \right>_F &=&
\left( -\Delta + \frac{\num{4.976e9}}{\rho^6} \right) E_h \,,
\end{eqnarray}
\endnumparts
and
\numparts
\begin{eqnarray}
\left< E(4P_{3/2}, \frakF_z=0) \right>_F &=& 
\left( \frac12 \, \Delta + \frac{\num{4.386e9}}{\rho^6} \right) E_h \,, \\
\left< E(4P_{3/2}, \frakF_z=\pm 1) \right>_F &=& 
\left( \frac37 \, \Delta + \frac{\num{3.546e9}}{\rho^6} \right) E_h \,, \\
\left< E(4P_{3/2}, \frakF_z=\pm 2) \right>_F &=& 
\left( \frac14 \, \Delta + \frac{\num{1.820e9}}{\rho^6} \right) E_h \,, \\
\left< E(4P_{3/2}, \frakF_z=\pm 3) \right>_F &=& 0 \; \,.
\end{eqnarray}
\endnumparts
Finally, as a function of $J$, complete 
averaging over $F$ and $\frakF_z$ leads to the results
\numparts
\begin{eqnarray}
\label{4P2S_4P12}
\left< E(4P_{1/2}) \right>_{F,\frakF_z} &=& 
\left( -\frac34 \, \Delta + \frac{\num{7.755e9}}{\rho^6} \right) E_h \,, \\
\label{4P2S_4P32}
\left< E(4P_{3/2}) \right>_{F,\frakF_z} &=& 
\left( \frac38 \, \Delta + \frac{\num{3.103e9}}{\rho^6} \right) E_h \,.
\end{eqnarray}
\endnumparts
Without hyperfine resolution, there are four $J = 3/2$ states
and two $J = 1/2$ states.
Hence, an additional average over the fine-structure levels
leads to a cancellation of the term proportional to $\Delta$,
but the $1/\rho^6$ energy shift remains as an overall 
repulsive interaction among $4P$--$2S$ atoms.

For the $4P_{1/2}$--$2S$ and $4P_{3/2}$--$2S$ systems,
the \vdw{} interactions are repulsive,
and we obtain large \vdw{} coefficients of order $10^9$ in atomic units
[see Eqs.~(\ref{4P2S_4P12}) and~(\ref{4P2S_4P32})].
The large coefficients mainly are due to the 
virtual $(4S;2P_{1/2})$ states, which have to be
added to the quasi-degenerate basis, as outlined above.

%
% Averaging the Cross Sections
%
\section{Atom--Molecule Interactions}
\label{sec5}

%
% General Considerations
%
\subsection{General Considerations}
\label{sec51}

As already anticipated, for atomic beam spectroscopy,
it becomes necessary 
to investigate the van der Waals 
$C_6$ coefficient for collisions of 
highly excited hydrogen atoms
(in $P$ states), with hydrogen molecules.
Anticipating the result, we come to the 
conclusion that $|C_6| \lesssim 20$ in atomic units, 
but the analysis becomes tricky because of 
some vibrational sublevels of the 
H$_2$ Lyman and Werner bands, which are energetically 
rather close to the atomic-hydrogen 
$1S$--$4P$ and $1S$--$6P$ transitions.

Because of the presence of energetically lower 
virtual states in the systems, it is instructive to
start with a general consideration, expressing the $C_6$ coefficient
in terms of oscillator strengths and energy differences, for the 
two atomic or molecular systems undergoing the collision.
In order to allow for a compact notation,
we here switch to
atomic units [$\epsilon_0 = 1/(4 \pi)$, $\hbar = 1$, $c = 1/\alpha$].
In the non-retardation regime,  
the interatomic interaction between 
any two electrically neutral atoms or molecules
$A$ and $B$ is given as~\cite{AdEtAl2017vdWi,JeEtAl2017vdWii}
\begin{eqnarray}\label{E:int:NR}
E_{\mathrm{AB}}(R) &=& 
\mathrm{Re} \frac{3\mathrm{i} }{2\,\pi R^6} \int_{-\infty}^\infty 
\mathrm{d}\omega\, \alpha_A (\omega)\; \alpha_B (\omega)\,,
\end{eqnarray}
where $R$ is the interatomic distance (in atomic units,
i.e., measured in Bohr radii), and
$\alpha_J(\omega)$ is the dynamic polarizability of the $J$th atom
($J = A,B$), while $\mathrm{Re}$ stands for the real part.
The dynamic polarizability $\alpha_J(\omega)$ for atom $J$ in 
the reference state $| m \rangle$ reads 
\begin{eqnarray}\label{E:int:NR1}
\alpha_{J}(\omega) &=&
\frac{1}{3} \sum_{n} \mkern -23.3mu \int 
\left[  \frac{ \left|\left< m \left| \vec{r}\right| n\right>\right|^2 }%
{E_{nm} - \omega- \mathrm{i}\epsilon}
+  \frac{\left|\left< m \left| \vec{r}\right| n\right>\right|^2}%
{E_{nm} + \omega- \mathrm{i}\epsilon}\right]
\nonumber\\
&=& \sum_{n} \mkern -23.3mu \int \frac{1}{2\, E_{nm}} 
\left[ \frac{2}{3} \,E_{nm}\, 
\frac{\left|\left< m \left| \vec{r}\right| n\right>\right|^2}
{E_{nm} -\omega- \mathrm{i}\epsilon}
+\frac{2}{3} \,E_{nm}\,  \frac{\left|\left< m \left| \vec{r}\right| n\right>\right|^2}
{E_{nm} +\omega- \mathrm{i}\epsilon}\right]
\nonumber\\
&=& \sum_{n} \mkern -23.3mu \int \frac{1}{2\, E_{nm}} \sum_{\pm}
\frac{f_{nm}}{E_{nm} \pm \omega- \mathrm{i}\epsilon}
= \sum_{n} \mkern -23.3mu \int \frac{ f_{nm}}{ E_{nm}} \,
\frac{E_{nm} }{\left(E_{nm} - \mathrm{i}\epsilon\right)^2 - \omega^2}\,.
\end{eqnarray}
Here, $E_{nm} = E_n - E_m$ is the transition energy between the state
$|n\rangle$ and the state $|m\rangle$,
while $f_{nm} = 2/3\, E_{nm} \left|\left< m \left| \vec{r}\right| n\right>\right|^2$ 
is the dipole oscillator strength,
for the dipole-allowed virtual transition $| m \rangle \to | n \rangle$.  Note
that one has to sum over the magnetic quantum numbers of the virtual 
state $| n \rangle$, but one averages over the magnetic quantum numbers of the
reference state $| m \rangle$.

As an example, we calculate the dipole oscillator strength of 
$4P$--$1S$ transition in atomic hydrogen. 
In the following discussion, $f_{n'\ell',n\ell}$ indicates the dipole oscillator strength
for $n\ell\rightarrow n'\ell'$ transition.
For a $1S\rightarrow 4P$ transition, 
the dipole oscillator strength in atomic unit reads
\begin{eqnarray}
f_{41,10} &=& 
\frac{2}{3}\, E_{40} \left|\left<10 \left| \vec{r}\right| 4 1\right>\right|^2
\nonumber\\[0.1133ex]
&=& \frac{1}{3}\,\left(1-\frac{1}{4^2}\right) 
\left|\int_0^\infty R_{10}(r) \, r^3 \, R_{41}(r)\, \mathrm{d} r \right|^2\,,
\end{eqnarray}
where the radial functions $R_{10}(r)$ and $ R_{41}(r)$ in atomic units are given by
\begin{eqnarray}
R_{10}(r) &=& 2\;\exp\left(-r\right)\,,\quad 
R_{41}(r) = \frac{1}{16} \sqrt{\frac{2}{5!}\,} \; r\;
\exp\left(-\frac{r}{4}\right)
\, L_{2}^{3}\left(\frac{r}{2}\right)\,.
\end{eqnarray}
The associated Laguerre polynomials are denoted
as $L^m_n(x)$.
The integral for the transition matrix element 
can be evaluated analytically as
\begin{eqnarray}
\int_0^\infty R_{10}(r) \, r^3 \, R_{41}(r)\, \mathrm{d} r 
= \sqrt{\frac{2}{5!}\,}\frac{2^{12}\times 3^2}{5^6}\,.
\end{eqnarray}
Consequently, the dipole oscillator strength for the
$1S\rightarrow 4P$ transition reads
\begin{eqnarray}
f_{41,10} &=& \frac{2^{18}\times 3^3}{5^{12}}\simeq 2.8991\times 10^{-2} \, {\rm a.u.},
\end{eqnarray}
which agrees with Ref.~\cite{WiFu2009}.
The dipole oscillator strength of the $4P\rightarrow 1S$ transitions is related 
to that of the $1S\rightarrow 4P$ transition as
\begin{eqnarray}
\label{OSS:4p1s}
f_{10,41} &=& - \frac{g_{_{10}}}{g_{_{41}}}f_{41,10} = - \frac{2^{18}\times 3^2}{5^{12}}
\simeq -9.66\times 10^{-3} \, {\rm a.u.},
\end{eqnarray}
where $g_{_{n\ell}} = 2\ell+1 $ is the statistical weight 
for the $\left| n\ell \right>$ state.
The result~(\ref{OSS:4p1s}) holds for both 
$4P_{1/2}$ as well as $4P_{3/2}$ states.

Using Eq.~(\ref{E:int:NR1}), the interaction energy given in Eq.~(\ref{E:int:NR})
can be written as follows,
in the limit $\epsilon\rightarrow 0^+$
[see Eq.~(1) of Ref.~\cite{DaMoPe1967}],
\begin{eqnarray}\label{E:int:NR3}
E_{\mathrm{AB}}(R)= -\mathrm{Re} \, \frac{3}{2\,R^6} \sum_{nn'} \mkern -23.3mu \int 
\frac{ f_{nm}^{(A)}\, f_{n'm'}^{(B)}}%
{ E_{nm}^{(A)}\, E_{n'm'}^{(B)} \;(E_{nm}^{(A)}+ E_{n'm'}^{(B)})}
= - \frac{C_6}{R^6} \,,
\end{eqnarray}
where the sum-integral denotes the summation over
the discrete virtual states, and the integral over the continuum.
Alternatively, the van der Waals $C_6$-coefficient reads,
in terms of oscillator strengths and transition energies,
\begin{eqnarray}\label{C6:coeff}
C_6= \mathrm{Re} \, \frac{3}{2} \sum_{nn'} \mkern -23.3mu \int 
\frac{ f_{nm}^{(A)}\, f_{n'm'}^{(B)}}%
{ E_{nm}^{(A)}\, E_{n'm'}^{(B)} \;(E_{nm}^{(A)}+ E_{n'm'}^{(B)})} \,.
\end{eqnarray}
One observes that, in view of the correct placement of the poles (infinitesimal
imaginary parts in the propagator denominators), 
the sum of the level energies $E_{nm}^{(A)}+ E_{n'm'}^{(B)}$ enters the 
expression for $C_6$ (not the sum of their absolute magnitude, as one 
could otherwise falsely conclude, if one inconsistently performs the Wick rotation 
without considering the possible presence of poles in the 
first quadrant of the complex $\omega$-plane). 
If $| m\rangle$ is an excited state, such as the excited $4P$ state of atomic 
hydrogen, and $| n\rangle$ is the ground state, then 
$E_{nm}^{(A)}$ is negative.
For virtual transitions from the ground $X$ state of the 
H$_2$ molecule to an excited $|n' \rangle= |B\rangle$ or
$|n'\rangle = |C\rangle$ state, $E_{n'm'}^{(B)}$ is positive.

In the case of quasi-degeneracy, one may have a 
situation of mutual compensation, i.e.,
$E_{nm}^{(A)}+ E_{n'm'}^{(B)} \approx 0$, and 
the $C_6$ coefficient can be enhanced in magnitude.
The energy difference $\left(E_{1S}-E_{4P}\right)$ is approximately 
equal to $-15/32$ atomic units (Hartree), 
so it is approximately equal to the negative half of the Hartree energy. 
Typical oscillator strengths in atomic hydrogen atoms are
of the order of unity. The quantity 
\begin{eqnarray}\label{DenoC6}
E_{nm}^{(A)}+ E_{n'm'}^{(B)}=E_{1S}^{H}-E_{4P}^{H}+ E_{n'}^{H_2} - E_{X}^{H_2} \,,
\end{eqnarray}
in Eq.~(\ref{E:int:NR3}) thus needs to be given special attention.
Here $X$ is the $X {}^1\Sigma^+_g$ ground state of H$_2$.

\begin{figure}[t!]
\begin{center}
\begin{minipage}{0.85\textwidth}
\begin{center}
\includegraphics [width=0.6\columnwidth]{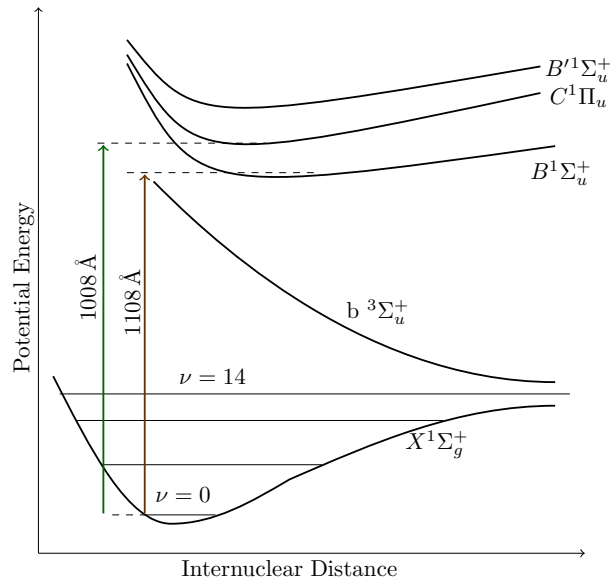}
\caption{\label{fig2} Schematic Born-Oppenheimer diagram for a hydrogen molecule
(not to scale).
The potential energy and the internuclear distance are given in arbitrary units. 
The ground state has 14 vibrational states which characterize the 
motion of the nuclei, 
while the excited $B{}^1 \Sigma_u^+$ and $C{}^1 \Pi_u$,
and $B'{}^1 \Sigma_u^+$  states
also harbor a number of vibrational sublevels.}
\end{center}
\end{minipage}
\end{center}
\end{figure}

%
% Molecular Spectrum
%
\subsection{Molecular Spectrum}
\label{sec52}

A short description of the molecular spectrum of the $H_2$ molecule is in order. 
Binding into $\Sigma$ states starts from two hydrogen 
atoms in the ground state
with orbital angular momenta
$L_{1,2} =0$ and electronic spin angular momenta $S_{1,2} =1/2$. 
As a result, the projection of the total angular
momentum onto the molecular axis is $\Lambda=L_1+L_2=0$,
and the total spin quantum number is $S=0,1$.
The first two excited states of $\Sigma$ symmetry,
above the molecular ground state  $ X ^1\Sigma_g^+$,
are $ B {}^1\Sigma_u^+$ and $ B' {}^1\Sigma_u^+$.
The spin-triplet $b {}^3\Sigma_u^+$ state is not bonding. 
Even if the $b {}^3\Sigma_u^+$ state were bonding,
we could ignore it because 
singlet to triplet transitions are forbidden 
by non-relativistic dipole selection rules~\cite{He1959}.
Electronic dipole transitions from the excited 
$B {}^1\Sigma_u^+$ and $C{}^1\Pi_u$ 
states of H$_2$ molecule (see Fig.~\ref{fig2}) to the ground state  $ X ^1\Sigma_g^+$  
were first observed by Lyman and Werner, and are 
therefore called the Lyman and Werner 
bands~\cite{Ly1906,We1926,HeHo1959,He1950}. 

The transition from the ground state $X {}^1\Sigma_g^+$ to the 
$B {}^1\Sigma_u^+$ state occurs at 1108\AA\; $\simeq$\;11.18991\,eV, 
while the $X {}^1\Sigma_g^+$--$C {}^1\Pi_u^+$ transition occurs 
at 1008\AA\;=\;12.30002\,eV (see Ref.~\cite{FiSoDr1996}). 
These figures exclude possible vibrational and rotational excitations.
The $1S$--$4P_{1/2}$ transition of atomic hydrogen occurs at 12.74851\,eV,
while the $1S$--$4P_{3/2}$ transition energy is 12.74852\,eV
(see Ref.~\cite{hdel}). Note that the 
$1S$--$4P_{1/2}$ and  $1S$--$4P_{3/2}$ transition energies 
differ only by the fine-structure (which,
in this case, enters at the seventh decimal).
Indeed, the fine-structure splitting 
is an effect of relative order $\alpha^2$ (see Ref.~\cite{ItZu1980}). 
The difference of the atomic $1S$--$4P_{1/2}$ transition energy 
to the $X$--$B$ and $X$--$C$ transitions is at least 0.4485\,eV,
provided no vibrational excitation occurs.
For comparison, the transition energies for the $2S$--$4P_{1/2}$  and  $3S$--$4P_{1/2}$ 
transitions~\cite{hdel} 
of atomic hydrogen are, respectively,  2.5497\,eV and 0.6610\,eV.

These considerations exclude vibrational and rotational excitations.
In general, the ro-vibrational energy of a molecule is given as 
\begin{eqnarray}\label{Eq:rovib}
E(\nu,J)&=& \left(\nu + \frac{1}{2} \right)\omega_e
- \left(\nu +\frac{1}{2}\right)^2 x_e\omega_e 
\nonumber\\
& & + B_\nu J (J+1) -D_J J^2 (J+1)^2\,,
\end{eqnarray}
where $\nu$ is the vibrational, and $J$ is the rotational quantum number.
Here,  $x_e\omega_e $ is the first-order
anharmonic correction to 
the harmonic oscillator approximation to molecular vibration.
The constant $B_\nu$,
\begin{eqnarray}
B_\nu= B_e - \alpha_e \left(\nu + \frac12 \right) \,,
\end{eqnarray}
is the rotational constant for a given vibrational state. 
Here, $B_e$ is the rotational constant in the equilibrium position, 
and $\alpha_e$ is the first-order anharmonicity
correction to  the rotational constant.
Finally, $D_J$ in Eq.~(\ref{Eq:rovib}) is the centrifugal distortion constant, 
several orders of magnitude smaller than $B_\nu$.  
To a first approximation,  we can assume that the molecular 
vibration is of harmonic oscillator type, and centrifugal distortions 
of the rotational levels
is negligible. More explicitly,
\begin{eqnarray}
\label{Eapprox}
E(\nu,J)\simeq \left(\nu +\frac{1}{2}\right)\omega_e+B_\nu \, J (J+1) \,.
\end{eqnarray}
Allowed ro-vibrational transitions have $\Delta J=0, \pm 1, \pm 2$. 
The $\Delta J=0$ transitions ($\Delta\nu\neq 0$) is the 
Raman $Q$-branch, while
the R-branch and P-branches correspond to the $\Delta J=+1$ and $\Delta J=-1$
transitions, and are relevant for pure rotational spectroscopy. 
For diatomic molecules, in Raman transitions, the selection rules 
imply that the allowed transitions have
$\Delta J=+2$ and $\Delta J=-2$ (Stokes and 
anti-Stokes lines, so-called S and O branches). 
Transitions with $|\Delta J|>2 $  are forbidden by  
selection rules~\cite{He1959}.
Here, we are neither concerned with pure rotational spectroscopy,
nor with Raman spectroscopy, but with the inclusion of the 
ro-vibrational transitions into the 
sum-over-states representation of the 
$C_6$ coefficient according to Eq.~(\ref{C6:coeff}).
In order to discern the allowed rotational transitions
from the $X$ to the $B$ and $C$ states,
one needs to observe that the $X$ state is 
{\rm gerade}, while $B$ and $C$ are {\em ungerade}.
For the molecular ground state, it is well known that, if the 
proton spins in H$_2$ are antiparallel (total proton
spin zero), then the spin wave function is antisymmetric
under particle (proton) interchange, 
so that the orbital proton wave function must be
symmetric under particle (proton) interchange,
resulting in even values for $J$ (para-hydrogen).
By contrast, if the proton spins in H$_2$ are parallel (total proton
spin one), then the spin wave function is symmetric
under particle (proton) interchange,
and the orbital proton wave function must be
anti-symmetric under particle (proton) interchange,
resulting in odd values for $J$ (ortho-hydrogen).
This holds because the ground-state two-electron wave function
is {\em gerade}, while the required proton wave function symmetry
is reversed for the $B$ and $C$ states, which are 
{\em ungerade} (see Ref.~\cite{DeKe1971}).
One can understand the symmetries most easily if
one considers the molecular wave function in 
the Born--Oppenheimer approximation~\cite{He1950}.

We have thus shown that the virtual transitions entering
the expression (\ref{C6:coeff}) have $\Delta J = \pm 1$ if the 
transition involves {\em gerade} and {\em ungerade} states
of the hydrogen molecule.
Let us try to analyze the frequency shift in a 
virtual transitions of H$_2$, due to the 
addition of a rotational quantum.
We anticipate that,
because of the small magnitude of the effect,
it is sufficient to study the frequency shift within 
a given manifold of rotational states,
specific to either the initial or the final state
of the virtual transition.
The energy differences for 
$J \to J+1$ transitions, within a given vibrational band, read as
\begin{eqnarray}
\Delta E(\nu,J) = E(\nu,J+1) - E(\nu,J)
\simeq 2B_\nu \,(J + 1) \,.
\end{eqnarray}
The difference between rotational lines in a vibrational band 
is thus $\Delta E(\nu,J+1)-\Delta E(\nu,J)=2B_\nu$,
which means  the ro-vibrational transition energies increase equally 
by an amount of $2B_\nu$ in both $\Delta J=\pm 1$. 
For the $X {}^1\Sigma_g^+$ state, the rotational $B_e$ and $\alpha_e$ 
constants are $B_e=60.853\, \mathrm{cm}^{-1}$ and 
$\alpha_e= 3.062\, \mathrm{cm}^{-1}$,
respectively (see Ref.~\cite{He1950}). Thus,
the $B_\nu$ coefficient of the $X {}^1\Sigma_g^+$ state for the 
vibrational ground state is $7.355 \times 10^{-3} \,\mathrm{eV}$.

%
% Possible Enhancement of the \vdw{} Coefficient
%
\subsection{Possible Enhancement of the \vdw{} Coefficient}
\label{sec53}

We have already stressed that the difference of the atomic 
$1S$--$4P$ transition to the 
$X$--$B$ and $X$--$C$ transitions of H$_2$ is at least 0.4485\,eV,
thus setting a lower limit for the magnitude of the propagator denominator
given in Eq.~(\ref{DenoC6}). 
Two effects could lead to an enhancement of $C_6$.
{\em (i)} One might assume that the ground-state hydrogen molecule 
enters the collision with atomic hydrogen, 
in a thermally excited rotational 
state, thus modifying the transition frequencies to 
virtual excited states of the molecule,
and {\em (ii)} potential virtual transitions from the $X$ ground state 
of H$_2$ to rotational sidebands of the vibrational levels $\nu =11$ 
of the $B$, and $\nu = 2$ of the $C$, state, could potentially
enhance $C_6$.

Let us try to address point {\em (i)}.
At a temperature of $T=5.8\,{\rm K}$,
which is relevant for the experiment~\cite{BeEtAl2017}
the thermal excitation energy is 
$k_B T = 4.998\times 10^{-4}\, \mathrm{eV}$. 
(In general, high-precision atomic-beam experiments
profit enormously from cryogenic beams.)
Equating the thermal excitation energy with the rotational energy, 
one can obtain an estimate for the typical rotational $J$ value
due to thermal excitation,  assuming a Boltzmann distribution, 
\begin{eqnarray}\label{Eq:J:Ro}
J\, (J+1)-\frac{4.998\times 10^{-4} \mathrm{eV}}{7.355 \times 10^{-3} \,\mathrm{eV}}=0
\Rightarrow J=0.06\,.
\end{eqnarray}
This implies that the thermal energy is insufficient to excite rotational levels, 
leaving the molecular ground state $X {}^1\Sigma_g^+$ of the system 
in the rotational ground state of the $\nu=0$ vibrational band.
Thus, we can safely assume that all collisions involving 
H$_2$ molecules start from the rotational ground state,
i.e., from a para-hydrogen state (after thermalization).

Having excluded thermal excitation of the 
ground state as a further source of a quasi-degeneracy 
of transitions in our system, we must now exclude point~{\em (ii)},
namely, the 
possibility of virtual transitions, from the rotational ground state
of the hydrogen molecule, to higher vibrational and rotational sublevels of the 
$B$ and $C$ states, which could otherwise drastically reduce
the energy difference with respect to the hydrogen $1S$--$4P$
transition, and decrease the magnitude of the quantity
$E_{nm}^{(A)}+ E_{n'm'}^{(B)}$ in Eq.~(\ref{DenoC6}).
We recall that the energy difference between the atomic $1S$--$4P$ transition and 
the $X$--$B$ molecular transition is 1.5589\,eV.
The $\nu=11$  vibrational sublevel of the $B {}^1\Sigma_u^+$
state of molecular hydrogen  has an energy of % 
102856.97 cm$^{-1}\simeq$ 12.7526\,eV (see Table I of Ref.~\cite{St1973}), 
which is closest to the $1S$--$4P$ transition of 12.7485\,eV,
among all vibrational levels
but higher in energy than the atomic hydrogen line,
so that the degeneracy cannot be reduced by adding
rotational quanta. 
For the $B$ transition, in order to address the possibility of
rotationally induced quasi-degeneracy, one should also note the
$\nu=10$ vibrational sublevel of the $B {}^1\Sigma_u^+$
state of molecular hydrogen  has an energy of % 
101864.90 cm$^{-1}\simeq$ 12.6296\,eV (see Table I of Ref.~\cite{St1973}).
On the other hand, 
we recall once more that the energy difference between the atomic $1S$--$4P$ 
transition and $X$--$C$  transition in molecular hydrogen is 0.4485\,eV. 
The $\nu=2$ vibrational sublevel of the $C {}^1\Pi_u$
state of molecular hydrogen  has an energy of % 
103628.662 cm$^{-1}\simeq$ 12.84830\,eV (see Table 5 of Ref.~\cite{PhEtAl2004}), 
which is very close to the $1S$--$4P$ transition of 12.7485\,eV
and energetically closest among the different vibrational 
levels. As an inspection shows,
it is also higher in energy than the atomic hydrogen line,
so that the degeneracy cannot be reduced by adding
rotational quanta.
For the $X$--$C$ transition, in order to address the possibility of
rotationally induced quasi-degeneracy, one should also note the
$\nu=1$ vibrational sublevel of the $C {}^1\Pi_u$
state of molecular hydrogen  has an energy of % 
101457.569 cm$^{-1}\simeq$ 12.5791\,eV (see Table 5 of Ref.~\cite{PhEtAl2004}).

One can argue as follows.
The rotational energy roughly follows $J(J+1)\approx J^2$, for large $J$
[see Eq.~(\ref{Eapprox})]. 
For the $X$--$B$ transition, to achieve quasi-degeneracy 
of about $1.189 \times 10^{-1}\, \mathrm{eV}$ 
with $B_\nu\sim 9.395 \times 10^{-4}\, \mathrm{eV}$,
we need $J^2\sim127\Rightarrow J\sim 11$.
Likewise,  for the $X$--$C$ transition, in order to achieve quasi-degeneracy 
by adding rotational excitation energy of the excited H$_2$ state
of about $1.694 \times 10^{-1}\, \mathrm{eV}$ 
with $B_\nu\sim 3.579\times 10^{-3}\, \mathrm{eV}$,
we need $J^2\sim 47 \Rightarrow J\sim 7$. 
By symmetry considerations, one can show that relevant 
rotational transitions in our system
need to  satisfy $\Delta J = \pm 1$.
Transitions with $\Delta J = +1$ bring the $X$--$B$ and  the $X$--$C$ transitions 
closer to the $1S$--$4P_{1/2}$ atomic transition only by 0.1\% and  1.5\%
respectively. With a forbidden transition
featuring $\Delta J = +2$, one can bring the $X$--$B$ and  
the $X$--$C$ transitions closer to the $1S$--$4P$ atomic transition 
only by the insignificant amounts of 0.3\% and  4.5\%, respectively. 
Effects due to higher multipoles, which could
potentially lead to ``even more forbidden'' transitions, 
are typically
suppressed by powers of $\alpha$~\cite{YaBaDaDr1996,LaDKJe2010pra},
with one power of $\alpha$ for each higher angular momentum involved.
For the very high required $\Delta J$ values, 
the contribution from the transitions which involve the ``highly 
forbidden $\Delta J$''
is thus numerically suppressed and can safely be neglected.

%
% Estimate of the \vdw{} Coefficient
%
\subsection{Estimate of the \vdw{} Coefficient}
\label{sec54}

The remaining task is 
to find the oscillator strength of  excitation
from the ground $X {}^1\Sigma_g^+$ molecular state to the $\nu=11$ vibrational
side band of the excited $B {}^1\Sigma_u^+$ molecular state,
and the same for the relevant $X$--$C$ transition.
The oscillator strength for the $\nu=11$ vibrational band 
of the Lyman band is given in Ref.~\cite{ChCoBr1992,AlDa1969} and reads 
and $f=1.74\times 10^{-2}\, {\rm a.u.}$,
while the oscillator strength for $\nu=2$ vibrational band  of the Werner
band is $f'=6.95 \times 10^{-2} \, {\rm a.u.}$
(see Refs.~\cite{ChCoBr1992,AlDa1969}). 
For comparison, slightly discrepant oscillator strengths
are given in Refs.~\cite{GeSc1969} and \cite{FaLe1974},
for the $\nu=2$ vibrational band 
of $C$, namely, $f'= 5.55\times 10^{-2}$  and  $f'= 6.42\times 10^{-2}$,
respectively. We here use the oscillator strength reported in Ref.~\cite{ChCoBr1992}
in our estimate.
The oscillator strength for the $4P$--$1S$
atomic hydrogen transition is $-9.66\times 10^{-3}$ [see Eq.~(\ref{OSS:4p1s})]. 
Consequently, the 
contribution of the virtual vibrational sublevels of the 
$B$ and $C$ states of H$_2$, which are  closest-in-energy to the 
$1S$-$4P$ transition in H, are given as 
\begin{eqnarray}\label{C6:coeffXB}
C_6(X;B)\approx  \frac{3}{2} \;
\frac{(-9.66 \times 10^{-3})\times 1.74 \times 10^{-2}}%
{ (-0.4685)\times 0.4686\times 1.499\times 10^{-4}}
= 7.661\; \mathrm{a.u.} \,,
\end{eqnarray}
\begin{eqnarray}\label{C6:coeffXC}
C_6(X;C)\approx  \frac{3}{2} \;
\frac{ (-9.66 \times 10^{-3})\times 6.95\times 10^{-2}}%
{ (-0.4685)\times 0.4722\times 3.667\times 10^{-3}}
= 1.241\; \mathrm{a.u.} \,.
\end{eqnarray}
The sum is $C_6(X;B)+C_6(X;C)$ is $\sim 8.901$ in atomic units. 
The energies in the  denominator 
of Eqs.~(\ref{C6:coeffXB}) and (\ref{C6:coeffXC}) are expressed in terms of the atomic 
unit of energy, namely, the Hartree energy $E_h = 27.2114 \, {\rm eV}$,
using the unit conversion of $1\,{\rm eV} = 0.0367493 E_h$.  	
As a last step, one needs to consider $X$--$B'$ transitions.
Neglecting rotational quanta,
the $X$--$B'$ transition is at 110529.47\,cm$^{-1}\simeq 13.704$\,eV 
(see Table~5 of~\cite{AbEtAl1994}),
while the $X$--$D$ transition is at 1129335.29\,cm$^{-1}\simeq 14.002$\,eV 
(see Table~7 of~\cite{AbEtAl1994}). These transition energies
exceed the ionization threshold of atomic hydrogen. 
Considering the $X$--$B'$  and  $X$--$D$ transitions 
in the H$_2$ molecule and the $4P$--$1S$ transition in atomic hydrogen, 
the propagator denominator (\ref{DenoC6}) becomes positive,
and, in magnitude, greater than the $4P$ atomic 
hydrogen binding energy.
Consequently, the contribution of the $B'$ and the $D$ states
to the van der Waals $C_6$ coefficient
in the \mbox{H($4P$)--H$_2$} interactions is opposite in sign 
to that of $B$ and $C$ molecular states;
numerically, it is small in magnitude
in comparison to $C_6(X;B)$ and $C_6(X;C)$. 
Because the involved virtual transition frequencies and oscillator 
strengths are independent 
of the hydrogen fine structure, to the order of the approximations 
made, the result is the same for both $4P_{1/2}$ and $4P_{3/2}$ reference states.
We can thus safely neglect the possibility of a dramatic enhancement of the 
$C_6$ coefficient in collisions of hydrogen molecules with $4P$ hydrogen atoms.
The total magnitude of the $C_6$ coefficient will be determined by
non-quasi-degenerate states, i.e., by a sum over the entire
bound and continuous spectrum of the hydrogen atoms and
molecules, as given by the general formula~(\ref{C6:coeff}).
Based on typical calculations available for other atomic
and molecular systems without quasi-degeneracies~\cite{DaMoPe1967},
we can thus conservatively estimate that
\begin{eqnarray}
\label{estimate1}
| C_6(4P \, {\rm H}; X {}^1 \Sigma^+_g \, {\rm H}_2) | \leq 20 \, {\rm a.u.} \,.
\end{eqnarray}

Let us now turn to the H--H$_2$ interaction for 
the planned $1S$--$6P$ experiment~\cite{UdHaPriv2017}. 
The $6P$--$1S$ transition energy
of about $13.22068 \,$eV~\cite{hdel}
is comparable to the $X$--$B(\nu=15)$ transition 
energy of 106534.3\, cm$^{-1}\simeq 13.2085\,$eV~\cite{St1973jcp}
and the $X$--$C(\nu=4)$ transition energy 
of 107580.936\,cm$^{-1}\simeq 13.3383\,$eV (see Ref.~\cite{PhEtAl2004}). 
The binding energy of the $6P$-level of atomic hydrogen 
is less than that of the $4P$-level.
We notice that the $X$--$B(\nu=15)$ transition 
energy is below the atomic $6P$--$1S$ transition energy (in absolute 
magnitude),
while the magnitude of the $X$--$C(\nu=4)$ transition energy exceeds
that of the atomic $6P$--$1S$ energy difference.
For the same reasons as given above for $4P$ interactions,
the $B'{}^1\Sigma_u^+$ and $D{}^1\Pi_u$ molecular levels 
lead to negligible contributions to the $C_6$
coefficient for \mbox{H($6P$)--H$_2$} interactions.
The oscillator strength of the 
$\nu=15$ vibrational level of the molecular $B$ state 
and the $\nu=4$ vibrational level of the molecular $C$ state are,
respectively, $7.94\times 10^{-3}$ and $3.87\times 10^{-2}$
in atomic units~\cite{ChCoBr1992}. 
The oscillator strength for the $6P_{j}$--$1S$ transition, where $j$ 
takes either $1/2$ or $3/2$, is $-2.60\times 10^{-3}$. 
As a result, the $C_6$ coefficient of the 
\mbox{H($6P$)--H$_2$}
interactions reads $C_6= 2\times (-0.293+0.147)\,{\rm a.u.}=-0.292 \, {\rm a.u.}$.
Just as for $4P$ hydrogen, 
the total magnitude of the $C_6$ coefficient will be determined by 
non-quasi-degenerate states.
molecules, as given by the general formula~(\ref{C6:coeff}).
Based on typical calculations available for other atomic
and molecular systems without quasi-degeneracies~\cite{DaMoPe1967},
we can thus conservatively estimate that
\begin{eqnarray}
\label{estimate2}
| C_6(6P \, {\rm H}; X {}^1 \Sigma^+_g \, {\rm H}_2) | \leq 20 \, {\rm a.u.} \,.
\end{eqnarray}
Both estimates~(\ref{estimate1}) and~(\ref{estimate2})
are smaller than the $C_6$ coefficients obtained
for atom-atom collisions, discussed in Secs.~\ref{sec3} 
and~\ref{sec4}.

%
% Conclusions
%
\section{Conclusions}
\label{sec6}

We have studied the \vdw{} interaction 
of excited $4P$ hydrogen atoms with 
ground-state $1S$ and metastable $2S$ atoms,
and with hydrogen molecules.
In order to obtain reliable estimates of the 
\vdw{} interaction coefficients, one needs 
to expand the states in a hyperfine-resolved 
basis, and consider all off-diagonal matrix elements 
of the \vdw{} interaction Hamiltonian, as outlined 
in Secs.~\ref{sec2A} and~\ref{sec2B}.
The explicit construction of the 
hyperfine-resolved states is discussed in Sec.~\ref{sec2C},
and the use of the Wigner--Eckhart theorem for the 
calculation of the matrix elements of the 
\vdw{} interaction is described in Sec.~\ref{sec2D}.

For the $4P$--$1S$ system, one needs to 
include both the $4P_{1/2}$ as well as the 
$4P_{3/2}$ states in the quasi-degenerate basis,
because the $4P$ fine-structure frequency is 
commensurate with the $1S$ hyperfine transition 
splitting (see Sec.~\ref{sec3A}).
The matrix elements of the total 
Hamiltonian involve the so-called hyperfine--fine--structure 
mixing term (see Sec.~\ref{sec3B}),
which couples the $4P_{1/2}(F=1)$ to the 
$4P_{3/2}(F=1)$ levels [see Eq.~(\ref{HFSFSmix})].

The explicit matrices of the total Hamiltonian~(\ref{H})
in the manifolds with $\frakF_z = 3,2,1$ are 
described in Secs.~\ref{sec3C}---\ref{sec3E}.
Final results are also indicated for the 
(otherwise excessively complex) manifold with $\frakF_z = 0$.
Due to mixing terms of first order in the
\vdw{} interaction
between degenerate states in the two-atom system,
the leading term in the \vdw{} energy,
upon rediagonalization of the Hamiltonian matrix,
is of order $1/R^3$ for the
$4P$--$1S$ interaction, but it averages out 
to zero over the magnetic projections.
The phenomenologically important second-order 
shifts of the energy levels are given in Sec.~\ref{sec3F},
with various averaging procedures illustrating the 
dependence of the shifts on the quantum numbers,
and the dependence of the repulsive or 
attractive character of the interaction on the 
hyperfine-resolved levels.

The same procedure is applied to the 
$4P$--$2S$ interaction in Sec.~\ref{sec4},
with the additional complication that 
virtual quasi-degenerate $(4S;2P_{1/2})$
also need to be included in the basis.
The treatment of the $4P$--$1S$ and 
$4P$--$2S$ long-range interactions reveals the 
presence of numerically large coefficients multiplying the 
$1/\rho^6$ interaction terms, due to the presence 
of quasi-degenerate levels.
The interaction remains nonretarded over all 
phenomenologically relevant distance scales.

For atom-molecule collisions, the analysis has been
carried out in Sec.~\ref{sec5}.
After some general considerations which 
illustrate the complications that can arise for 
excited states (see Sec.~\ref{sec51}),
we briefly discuss the molecular spectrum (Sec.~\ref{sec52}),
before discussing possible enhancement mechanisms
for the \vdw{} coefficient, which can be of thermal 
and other origin (see Sec.~\ref{sec53}).
A numerical estimate of the coefficient is 
performed in Sec.~\ref{sec54},
with the result that the drastic enhancement that 
we see in atom-atom collisions, is in fact 
absent for atom-molecular interactions.
This observation is of high relevance to the 
analysis of experiments.

\section*{Acknowledgments}

The authors acknowledge helpful conversations with 
Professor T W H\"{a}nsch, 
V~Debierre, and Th Udem.
This research has been supported by the 
National Science Foundation (Grants PHY--1710856 and
CHE-1566246).
N~K~acknowledges support from DFG-RFBR grants (HA 1457/12-1 and 17-52-12016).
A~M~acknowledges support from the 
Deutsche Forschungsgemeinschaft (DFG grant MA 7628/1-1).

\hspace*{1cm}


\noindent
{\bf References}

\begin{thebibliography}{10}

\bibitem{Ch1972}
M.~I. Chibisov, {\em \relax{Dispersion Interaction of Neutral Atoms}},  Opt.
  Spectrosc. {\bf 32},  1--3  (1972).

\bibitem{DeYo1973}
W.~J. Deal and R.~H. Young, {\em \relax{Long--Range Dispersion Interactions
  Involving Excited Atoms; the H(1s)---H(2s) Interaction}},  Int. J. Quantum
  Chem. {\bf 7},  877--892  (1973).

\bibitem{AdEtAl2017vdWi}
C.~M. Adhikari, V. Debierre, A. Matveev, N. Kolachevsky, and U.~D. Jentschura,
  {\em \relax{Long-range interactions of hydrogen atoms in excited states.~I.
  $2S$--$1S$ interactions and Dirac--$\delta$ perturbations}},  Phys. Rev. A
  {\bf 95},  022703  (2017).

\bibitem{JeEtAl2017vdWii}
U.~D. Jentschura, V. Debierre, C.~M. Adhikari, A. Matveev, and N. Kolachevsky,
  {\em \relax{Long-range interactions of excited hydrogen atoms. II.
  Hyperfine-resolved $2S$--$2S$ system}},  Phys. Rev. A {\bf 95},  022704
  (2017).

\bibitem{PoTh1995}
E.~A. Power and T. Thirunamachandran, {\em \relax{Dispersion forces between
  molecules with one or both molecules excited}},  Phys. Rev. A {\bf 51},
  3660--3666  (1995).

\bibitem{SaBuWeDu2006}
H. Safari, S.~Y. Buhmann, D.-G. Welsch, and H.~T. Dung, {\em
  \relax{Body-assisted van der Waals interaction between two atoms}},  Phys.
  Rev. A {\bf 74},  042101  (2006).

\bibitem{SaKa2015}
H. Safari and M.~R. Karimpour, {\em \relax{Body-Assisted van der Waals
  Interaction between Excited Atoms}},  Phys. Rev. Lett. {\bf 114},  013201
  (2015).

\bibitem{Be2015}
P.~R. Berman, {\em \relax{Interaction energy of nonidentical atoms}},  Phys.
  Rev. A {\bf 91},  042127  (2015).

\bibitem{MiRa2015}
P.~W. Milonni and S.~M.~H. Rafsanjani, {\em \relax{Distance dependence of
  two-atom dipole interactions with one atom in an excited state}},  Phys. Rev.
  A {\bf 92},  062711  (2015).

\bibitem{DoGuLa2015}
M. Donaire, R. Gu\'{e}rout, and A. Lambrecht, {\em \relax{Quasiresonant van der
  Waals Interaction between Nonidentical Atoms}},  Phys. Rev. Lett. {\bf 115},
  033201  (2015).

\bibitem{Do2016}
M. Donaire, {\em \relax{Two-atom interaction energies with one atom in an
  excited state: van der Waals potentials versus level shifts}},  Phys. Rev. A
  {\bf 93},  052706  (2016).

\bibitem{JeAdDe2017prl}
U.~D. Jentschura, C.~M. Adhikari, and V. Debierre, {\em \relax{Virtual Resonant
  Emission and Long--Range Tails in van der Waals Interactions of Excited
  States: QED Treatment and Applications}},  Phys. Rev. Lett. {\bf 118},
  123001  (2017).

\bibitem{BeEtAl2017}
A. Beyer, L. Maisenbacher, A. Matveev, R. Pohl, K. Khabarova, A. Grinin, T.
  Lamour, D.~C. Yosta, T.~W. H\"{a}nsch, N. Kolachevsky, and T. Udem, {\em
  \relax{The Rydberg constant and proton size from atomic hydrogen}},  Science
  {\bf 358},  79--85  (2017).

\bibitem{Wo1999}
S. Wolfram, {\em \relax{The Mathematica Book}}, 4 ed. (Cambridge University
  Press, Cambridge, UK, 1999).

\bibitem{JoEtAl2002}
S. Jonsell, A. Saenz, P. Froelich, R.~C. Forrey, R. C\^{o}t\'{e}, and A.
  Dalgarno, {\em \relax{Long-range interactions between two 2s excited hydrogen
  atoms}},  Phys. Rev. A {\bf 65},  042501  (2002).

\bibitem{MaEtAl2018jpb2}
A. Matveev, N. Kolachevsky, C.~M. Adhikari, and U.~D. Jentschura, {\em Pressure
  Shifts in High--Precision Hydrogen Spectroscopy. II. Impact Approximation and
  Monte--Carlo Simulations}, submitted (2018).

\bibitem{PoPriv2017}
R. Pohl, private communication (2017).

\bibitem{UdPriv2017}
T. Udem, private communication (2017).

\bibitem{JeAd2017atoms}
U.~D. Jentschura and C.~M. Adhikari, {\em \relax{Long--Range Interactions for
  Hydrogen: $6P$--$1S$ and $6P$--$2S$ Systems}},  Atoms {\bf 5},  48  (2017).

\bibitem{BeHiBo1995}
D.~J. Berkeland, E.~A. Hinds, and M.~G. Boshier, {\em \relax{Precise Optical
  Measurement of Lamb Shifts in Atomic Hydrogen}},  Phys. Rev. Lett. {\bf 75},
  2470--2473  (1995).

\bibitem{MoNeTa2016}
P.~J. Mohr, D.~B. Newell, and B.~N. Taylor, {\em \relax{CODATA Recommended
  Values of the Fundamental Physical Constants: 2014}},  Rev. Mod. Phys. {\bf
  88},  035009  (2016).

\bibitem{AdDeJe2017aphb}
C.~M. Adhikari, V. Debierre, and U.~D. Jentschura, {\em \relax{Adjacency graphs
  and long-range interactions of atoms in quasi-degenerate states: applied
  graph theory}},  Appl. Phys. B {\bf 123},  1  (2017).

\bibitem{Sa2016}
A. Salam, {\em \relax{Non-Relativistic QED Theory of the van der Waals
  Dispersion Interaction}} (Springer, Cham, Switzerland, 2016).

\bibitem{Ad2017phd}
C.~M. Adhikari, Ph.D. thesis, Missouri University of Science and Technology,
  Rolla, MO, 2017 (unpublished).

\bibitem{ItZu1980}
C. Itzykson and J.~B. Zuber, {\em \relax{Quantum Field Theory}} (McGraw-Hill,
  New York, 1980).

\bibitem{Ed1957}
A.~R. Edmonds, {\em \relax{Angular Momentum in Quantum Mechanics}} (Princeton
  University Press, Princeton, New Jersey, 1957).

\bibitem{BrSa1994}
D.~M. Brinks and G.~R. Satchler, {\em \relax{Angular Momentum}} (Oxford
  University Press, Oxford, 1994).

\bibitem{LaLi1958vol3}
L.~D. Landau and E.~M. Lifshitz, {\em \relax{Quantum Mechanics, Volume 3 of the
  Course on Theoretical Physics}} (Pergamon Press, Oxford, UK, 1958).

\bibitem{BeSa1957}
H.~A. Bethe and E.~E. Salpeter, {\em \relax{Quantum Mechanics of One- and
  Two-Electron Atoms}} (Springer, Berlin, 1957).

\bibitem{HoHo2016}
M. Horbatsch and E.~A. Hessels, {\em \relax{Tabulation of the bound-state
  energies of atomic hydrogen}},  Phys. Rev. A {\bf 93},  022513  (2016).

\bibitem{hdel}
For an interactive database of hydrogen and deuterium transition frequencies,
  see the URL http://physics.nist.gov/hdel.

\bibitem{KoEtAl2009}
N. Kolachevsky, A. Matveev, J. Alnis, C.~G. Parthey, S.~G. Karshenboim, and
  T.~W. H\"{a}nsch, {\em \relax{Measurement of the $2S$ Hyperfine Interval in
  Atomic Hydrogen}},  Phys. Rev. Lett. {\bf 102},  213002  (2009).

\bibitem{LuPi1981}
S.~R. Lundeen and F.~M. Pipkin, {\em \relax{Measurement of the Lamb Shift in
  Hydrogen, $n=2$}},  Phys. Rev. Lett. {\bf 46},  232--235  (1981).

\bibitem{Pa1996mu}
K. Pachucki, {\em \relax{Theory of the Lamb shift in muonic hydrogen}},  Phys.
  Rev. A {\bf 53},  2092--2100  (1996).

\bibitem{Ed1974ang}
A.~R. Edmonds, {\em \relax{Angular Momentum in Quantum Mechanics}} (Princeton
  University Press, Princeton, New Jersey, 1974).

\bibitem{WiFu2009}
W.~L. Wiese and J.~R. Fuhr, {\em \relax{Accurate Atomic Transition
  Probabilities for Hydrogen, Helium, and Lithium}},  J. Phys. Chem. Ref. Data
  {\bf 38},  565--720  (2009).

\bibitem{DaMoPe1967}
A. Dalgarno, I.~H. Morrison, and R.~M. Pengelly, {\em \relax{Long-Range
  Interactions Between Atoms and Molecules}},  Int. J. Quantum Chem. {\bf 1},
  161--167  (1967).

\bibitem{He1959}
G. Herzberg, {\em \relax{Spectra of diatomic molecules}} (Van Nostrand
  Reinhold, Princeton, MA, 1959).

\bibitem{Ly1906}
T. Lyman, {\em \relax{The Spectrum of Hydrogen in the Region of Extremely Short
  Wave--Length}},  Astrophys. J. {\bf 23},  181--210  (1906).

\bibitem{We1926}
S. Werner, {\em \relax{Hydrogen Bands in the Ultra--Violet Lyman Region}},
  Proc. Roy. Soc. London, Ser. A {\bf 113},  107--117  (1926).

\bibitem{HeHo1959}
G. Herzberg and L.~L. Howe, {\em \relax{The Lyman bands of molecular
  hydrogen}},  Can. J. Phys. {\bf 37},  636--659  (1959).

\bibitem{He1950}
W. Heitler, {\em \relax{Quantum Theory of Radiation}} (Oxford University Press,
  New York, 1950).

\bibitem{FiSoDr1996}
G.~B. Field, W.~B. Somerville, and K. Dressler, {\em \relax{Hydrogen Molecules
  in Astronomy}},  Ann. Rev. Astron. Astrophysics {\bf 4},  207  (1966).

\bibitem{DeKe1971}
M.~J.~S. Dewar and J. Kelemen, {\em \relax{LCAO MO Theory Illustrated by Its
  Application to H$_2$}},  J. Chem. Edu. {\bf 48},  494--501  (1971).

\bibitem{St1973}
A.~F. Starace, {\em Comment on ``Length and Velocity Formulas in Approximate
  Oscillator--Strength Calculations''},  Phys. Rev. A {\bf 8},  1141--1142
  (1973).

\bibitem{PhEtAl2004}
J. Philip, J.~P. Sprengers, T. Pielage, C.~A. de~Lange, W. Ubachs, and E.
  Reinhold, {\em \relax{Highly accurate transition frequencies in the H 2 Lyman
  and Werner absorption bands}},  Can. J. Chem. {\bf 82},  713--722  (2004).

\bibitem{YaBaDaDr1996}
Z.~C. Yan, J.~F. Babb, A. Dalgarno, and G.~W.~F. Drake, {\em Variational
  calculations of dispersion coefficients for interactions among H, He, and Li
  atoms},  Phys. Rev. A {\bf 54},  2824--2833  (1996).

\bibitem{LaDKJe2010pra}
G. \L{}ach, M. DeKieviet, and U.~D. Jentschura, {\em \relax{Multipole Effects
  in Atom--Surface Interactions: A Theoretical Study with an Application to
  He--$\alpha$-quartz}},  Phys. Rev. A {\bf 81},  052507  (2010).

\bibitem{ChCoBr1992}
W.~F. Chan, G. Cooper, and C.~E. Brion, {\em \relax{Absolute optical oscillator
  strengths (11-20\,eV) and transition moments for the photoabsorption of
  molecular hydrogen in the Lyman and Werner bands}},  Chem. Phys. {\bf 168},
  375--388  (1992).

\bibitem{AlDa1969}
A.~C. Allison and A. Dalgarno, {\em \relax{Band oscillator strengths and
  transition probabilities for the Lyman and Werner systems of H$_2$, HD, and
  D$_2$}},  At. Data Nucl. Data Tables {\bf 1},  289--304  (1969).

\bibitem{GeSc1969}
J. Geiger and H. Schmoranzer, {\em \relax{Electronic and vibrational transition
  probabilities of isotopic hydrogen molecules H$_2$, HD, and D$_2$ based on
  electron energy loss spectra}},  J. Mol. Spect. {\bf 32},  39--53  (1969).

\bibitem{FaLe1974}
W. Fabian and B.~R. Lewis, {\em \relax{Experimentally determined oscillator
  strengths for molecular hydrogen-I. The Lyman and Werner bands above
  900\AA}},  J. Quant. Spectry. Rad. Transfer {\bf 14},  523--535  (1974).

\bibitem{AbEtAl1994}
H. Abgrall, E. Roueff, F. Launay, and J.-Y. Roncin, {\em \relax{The $B' \, {}^1
  \Sigma_u^+ \to X \, {}^1 \Sigma_g^+$ and $D \, {}^1 \Pi_u \to X \, {}^1
  \Sigma_g^+$ band systems of molecular hydrogen}},  Can. J. Phys. {\bf 72},
  856--865  (1994).

\bibitem{UdHaPriv2017}
T. Udem and T.~W. H\"{a}nsch, private communication (2017).

\bibitem{St1973jcp}
W.~C. Stwalley, {\em \relax{Potential energy curve of the $B\,{}^1 \Sigma^+_u$
  state of H$_2^*$}},  J. Chem. Phys. {\bf 58},  536--540  (1973).

\end{thebibliography}
\end{document}